\documentclass[fleqn,usenatbib,twocolumn,twocolappendix]{openjournal}

\usepackage{newtxtext,newtxmath}

\usepackage[T1]{fontenc}
\usepackage{ae,aecompl}

\usepackage{graphicx}	
\usepackage{epstopdf}
\usepackage{amsmath}	
\usepackage{amssymb}	
\usepackage{float}
\usepackage{multirow,array}
\usepackage{soul}
\usepackage[usenames, dvipsnames]{xcolor}  
\usepackage[colorlinks=true, allcolors=blue]{hyperref}

\usepackage{comment}
\usepackage[caption=false]{subfig}

\usepackage{cjhebrew}
\usepackage{fontawesome}
\usepackage[T1]{fontenc}
\newfam\hebfam
\font\tmp=rcjhbltx at10pt \textfont\hebfam=\tmp
\font\tmp=rcjhbltx at7pt  \scriptfont\hebfam=\tmp
\font\tmp=rcjhbltx at5pt  \scriptscriptfont\hebfam=\tmp
\edef\declfam{\ifcase\hebfam 
     0\or1\or2\or3\or4\or5\or6\or7\or8\or9\or A\or B\or C\or D\or E\or F\fi}
\mathchardef\tav   = "0\declfam 74

\newcommand{\orcidauthor}[3]{\author{\href{http://orcid.org/#1}{#2$^{#3}$}}}

\begin{document}

\title[\textsc{DeepDISC}: JWST photo-z]{Photometric Redshifts in JWST Deep Fields: A Pixel-Based Alternative with \textsc{DeepDISC}}

\orcidauthor{0009-0005-7923-054X}{Grant Merz}{1,*}
\orcidauthor{0000-0001-5105-2837}{Ming-Yang Zhuang}{1}
\orcidauthor{0000-0002-1605-915X}{Junyao Li}{1}
\orcidauthor{0000-0002-6893-3742}{Qian Yang}{2}
\orcidauthor{0000-0003-1659-7035}{Yue Shen}{1,3}
\orcidauthor{0000-0003-0049-5210}{Xin Liu}{1,3,4}
\orcidauthor{0000-0002-2495-3514}{John Franklin Crenshaw}{5,6,7}

\thanks{$^*$ Corresponding Author: \href{mailto:gmerz3@illinois.edu}{gmerz3@illinois.edu}.}


\affiliation{$^{1}$Department of Astronomy, University of Illinois at Urbana-Champaign, 1002 West Green Street, Urbana, IL 61801, USA}
\affiliation{$^{2}$ Center for Astrophysics | Harvard \& Smithsonian, 60 Garden St, Cambridge, MA 02138, USA}
\affiliation{$^{3}$National Center for Supercomputing Applications, University of Illinois at Urbana-Champaign, 1205 West Clark Street, Urbana, IL 61801, USA}
\affiliation{$^{4}$Center for Artificial Intelligence Innovation, University of Illinois at Urbana-Champaign, 1205 West Clark Street, Urbana, IL 61801, USA}
\affiliation{$^{5}$ Kavli Institute of Particle Astrophysics and Cosmology, P. O. Box 2450, Stanford, CA 94305}
\affiliation{$^{6}$ Department of Physics, Stanford University, 382 Via Pueblo Mall, Stanford, CA 94305}
\affiliation{$^{7}$ SLAC National Accelerator Laboratory, 2575 Sand Hill Road, Menlo Park, CA 94025}





\begin{abstract}
Photo-z algorithms that utilize SED template fitting have matured, and are widely adopted for use on high-redshift near-infrared data that provides a unique window into the early universe.  Alternative photo-z methods have been developed, largely within the context of low-redshift optical surveys.  Machine learning based approaches have gained footing in this regime, including those that utilize raw pixel information instead of aperture photometry. However, the efficacy of image-based algorithms on high-redshift, near-infrared data remains underexplored.  Here, we test the performance of Detection, Instance Segmentation and Classification with Deep Learning (\textsc{DeepDISC}) on photometric redshift estimation with NIRCam images from the JWST Advanced Deep Extragalactic Survey (JADES) program. \textsc{DeepDISC} is designed to produce probabilistic photometric redshift estimates directly from images, after detecting and deblending sources in a scene. Using NIRCam-only images and a compiled catalog of spectroscopic redshifts, we show that \textsc{DeepDISC} produces reliable photo-zs and uncertainties comparable to those estimated from template fitting using HST+JWST filters; \textsc{DeepDISC} even outperforms template fitting (lower scatter/fewer outliers) when the input photometric filters are matched. Compared with template fitting, \textsc{DeepDISC} does not require measured photometry from images, and can produce a catalog of 94000 photo-zs in ~4 minutes on a single NVIDIA A40 GPU. While current spectroscopic training samples are small and incomplete in color-magnitude space, this work demonstrates the potential of \textsc{DeepDISC} for increasingly larger image volumes and spectroscopic samples from ongoing and future programs.  We discuss the impact of the training data on applications to broader samples and produce a catalog of photo-zs for all JADES DR2 photometric sources in the GOODS-S field, with quality flags indicating caveats.

\end{abstract}


\section{Introduction}

\label{sec:intro}

Redshift measurements are a crucial data product for a multitude of astronomical surveys. While spectroscopy provides the most accurate method of measuring redshifts, it is time consuming and expensive, and is largely limited to bright photometric samples. Wide-area surveys such as the Legacy Survey of Space and Time \citep{LSST09}, Euclid \citep{EuclidQ1} and Roman \citep{RomanOverview}, will produce massive datasets, making complete spectroscopic catalogs impossible.  At the same time, the James Webb Space Telescope (JWST) is now revealing unprecedented numbers of galaxies at the highest redshifts in relatively narrow but ultradeep fields, e.g., the JWST Advanced Deep Extragalactic Survey \citep[JADES,][]{JADESDR1}, the Cosmic Evolution Early Release Science Survey \citep[CEERS,][]{CEERS}, and other JWST surveys. Even with JWST's powerful spectrographs, obtaining spectroscopic redshifts for the sheer abundance of faint, high-redshift galaxies is not possible.  

In lieu of spectroscopy, photometry can be used to estimate redshifts, since spectral features propagate into flux and color measurements and are thus strongly correlated with redshift. Due to the relatively coarse spectral energy distribution (SED) information provided by photometry, photometric redshifts, or photo-zs, are inherently less accurate and less precise than spectroscopic redshifts and face challenges due to degeneracies in the observed photometry and underlying redshift.  Different combinations of redshift and galaxy SED can produce very similar colors, leading to ambiguities and occasional catastrophic failures in photo-z estimation \citep{Massarotti01, Rieke23}. This issue can be exacerbated for JWST data in the early Universe: for example, a dusty galaxy at intermediate redshift can mimic the colors of a pristine galaxy at high redshift if only observed in infrared bands. Indeed, early JWST photometric studies uncovered numerous candidate galaxies at $z\gtrsim 10$-15 \citep{Naidu22,Donnan23}, some of which were later re-identified as lower-redshift interlopers after spectroscopic follow-up \citep{AH23}. Mitigating such outliers requires robust photo-z methodologies and an understanding of the limitations of the data. To quantify uncertainties, photo-z algorithms may produce a full redshift probability density function (PDF) for each object's photo-z rather than a single best value. Utilizing the entire PDF allows a more rigorous propagation of redshift uncertainties in scientific analyses. Photo-z PDFs are now standard in large photometric surveys \citep[e.g.,][]{Meyers09, DES16, DES21, HSC23} and will be a required data product for upcoming missions \citep[e.g.,][]{LSST_SRD}. JWST extragalactic programs likewise benefit from photo-z PDFs—for instance, to select high-z candidates for spectroscopy or to statistically constrain the redshift distribution of faint sources that cannot all be individually confirmed.

Techniques for estimating photometric redshifts generally fall into two broad categories: template SED-fitting approaches and machine learning approaches (see \cite{Salvato19,Schmidt20} for reviews). Template-based methods involve computing the goodness-of-fit of the observed photometry given a library of spectral templates (empirical or model SEDs) as a function of redshift. Examples include the widely used code \textsc{\texttt{EAZY}} \citep{Brammer08}, which was recently applied to derive photo-z for JWST observations in the JADES survey \citep{Rieke23,Hainline24}. Template fitting can incorporate priors on galaxy types or luminosities and naturally provides a redshift PDF (via $\chi^2$ or Bayesian likelihood). However, it requires accurate photometric calibration and error estimates, and its performance depends on the completeness of the template library.  For instance,  galaxies at high-z may have bluer rest-frame UV colors than typical templates and emission lines can pose further challenges if not represented \citep{Hainline24}. In contrast, machine learning (ML) methods attempt to learn the mapping from photometry to redshift by using a training set of objects with known redshifts, e.g., \cite{CarrascoKind13,FZB1}. Hybrid approaches that combine aspects of template fitting and machine learning have also been explored \citep{Tanigawa24,Crenshaw20}. 

Recently, there is growing interest in utilizing the images themselves as input to photo-z algorithms rather than precomputed photometric features. This image-based approach leverages the full information of pixels, including morphological characteristics such as object shape, size, surface brightness profile, and color gradients across the galaxy. High-resolution imaging from space telescopes like JWST offers an opportunity to exploit these morphological features, which may correlate with redshift (for instance, \cite{Huertas-Company23} find the fraction of JWST CEERS galaxies with ``irregular" morphological class to increase with redshift). A number of studies using seeing-limited images have demonstrated that deep learning models can extract informative features from images that improve photo-z estimates \citep{DIsanto18, Pasquet19, Hayat21, Dey22, Schuldt21, Treyer24, Ait-Ouahmed24}. Convolutional neural networks (CNNs) and other neural network architectures can non-linearly combine pixel information in multi-band image cutouts, often outperforming traditional SED-fitting or shallow ML algorithms given comparable data. Furthermore, advanced representation-learning techniques such as contrastive learning have been applied to astronomical images to pretrain feature extractors, which can then be fine-tuned for photometric redshift regression \citep{Hayat21,EuclidQ1AstroPT}. These studies suggest that image-based photo-z approaches are a promising complement to classical methods, particularly in low redshift regimes.  However, as we continue to leverage JWST's capabilities to observe the high-redshift universe, characterizing the feasibility of image-based photometric redshift models for these data is a worthwhile endeavor.

Despite their advantages, image-based and machine learning photo-z methods face several challenges, especially in the context of deep JWST surveys. First, one practical concern is object type contamination: most ML algorithms are developed for galaxies and thus assume that the input data are free of stars or QSOs, which can have very different SEDs. In wide-field surveys, a vast number of faint stars can masquerade as compact high-z galaxies in poor seeing, requiring robust star/galaxy separation \citep{Bosch18}. JWST deep fields are generally at high Galactic latitude and have very high spatial resolution, so stellar contamination is less severe, but some faint brown dwarfs or cool stars could still enter high-z galaxy samples if one relies purely on photometric criteria.  Many ML-based photo-z studies mitigate this by training on a pre-classified galaxy catalog. Notably, \cite{DIsanto18} demonstrate that a CNN could still yield accurate photo-zs on mixed-source data (stars, galaxies, and QSOs) without significant degradation from the presence of different object classes, suggesting that explicit pre-classification might be unnecessary if the model is sufficiently robust. \cite{Merz25} find that stellar contamination during training does not significantly affect image-based photo-z estimates for simulated LSST data. This question has not yet been fully tested on JWST data, an area for future exploration as more spectroscopic redshifts become available. 

Second, image-based methods take raw or minimally processed images as input; if two galaxies are blended, i.e., overlapping, or if there is simply a close neighboring source, a neural network might inadvertently use light from both in estimating a single redshift. Some authors have introduced techniques to mitigate this, such as masking neighboring objects in the input images \citep{Schuldt21} or explicitly modeling blending in template fitting techniques \citep{Jones19}. While blending is a low-order systematic in JWST data due its high resolution and $< 0\farcs1$ point-spread function, it is important to ensure that image-based methods correctly learn which pixels in an image correspond to a given source's redshift. 

Third, machine learning methods are limited in applicability outside of the feature space spanned by their training data.  Out-of-distribution estimation remains a problem, even for wide-field surveys with large training samples of spectroscopic data \citep{Moskowitz24,Zhang25,Desprez20, TSC24}.  Mitigating this effect is an ongoing effort and its implications will affect the results presented here.

In this work, we investigate photometric redshift estimation in the context of JWST NIRCam imaging by applying Detection, Instance Segmentation and Classification with Deep Learning:\textsc{DeepDISC photo-z}, an image-based deep learning framework for joint object detection/deblending, classification, and photo-z prediction. This builds on the method developed in \cite{Burke_deblending_2019} and the  work of \cite{Merz2023,Merz25} who develop \textsc{DeepDISC} and its photo-z extension.  Here, we extend \textsc{DeepDISC} to perform probabilistic redshift inference directly from multi-band JWST image cutouts, effectively bypassing the need for conventional photometric catalogs. By operating on the pixel level, our approach circumvents potential systematics from source extraction and aperture photometry, while simultaneously handling deblending. The network outputs a redshift probability distribution for each detected source, allowing us to capture uncertainties and degeneracies in the photo-z prediction. We train and evaluate our model using a combination of simulated and real JWST data. On the simulation side, we use mock JWST NIRCam images from the JAGUAR catalogs \citep{JAGUAR} to benchmark performance under controlled, known truth conditions. For real data, we utilize imaging from JADES in the Great Observatories Origins Deep Survey-South (GOODS-S) field, which provides deep 9-band NIRCam observations and a wealth of spectroscopic redshifts from early JWST/NIRSpec campaigns \citep{JADESDR1,JADESDR2}. This enables us to directly compare \textsc{DeepDISC} photo-z results to those from traditional SED-fitting on JWST photometry using \textsc{\texttt{EAZY}} and to assess the impact of various factors such as the number of filters available or training sample size. We also analyze failure modes, for example identifying regimes where our models might under-perform due to sparse training coverage or non-detections. Our study thus places this new image-based JWST photo-z technique in the broader context of existing methods, highlighting its strengths and current limitations.

In \S\ref{sec:data} we describe the JWST datasets used in this work, including the JADES imaging and spectroscopic catalogs and the simulated data generation. In \S\ref{sec:methods} we detail the \textsc{DeepDISC} photo-z model architecture and training methodology. In \S\ref{sec:results}, we present the results of our photometric redshift estimation on both the simulated and real JWST samples, and compare the performance to \texttt{EAZY}. In \S\ref{sec:discussion} we explore the dependence of the model performance on observational conditions, describe the application of our model to the entire JADES DR2 GOODS-S sample, and discuss the advantages and challenges of the image-based approach for JWST. Finally, we summarize our findings and conclude in \S\ref{sec:conclusions}. Our results demonstrate the viability of deep learning-based photo-z estimation for JWST and lay the groundwork for future applications to larger JWST surveys and forthcoming missions.

\section{Data}
\label{sec:data}

We use image mosaic and catalog data of the GOODS-S field from the second JADES data release \citep{JADESDR2}, hereafter referred to as JADES DR2 GOODS-S.  We use the 9 NIRCam filters corresponding to the first data release: F090W, F115W, F150W, F200W, F277W, F335M, F356W, F410M, and F444W.  Image data consists of observations from JWST Program 1210 and 1180 in Cycle 1, and Program 3215 in Cycle 2.   
Images are calibrated following the pipeline described in \cite{JADESDR1}. A detection image is created from stacking the long wavelength channel images, and \texttt{Photutils} \citep{photutils} is used for object detection, segmentation, and photometry. Imaging from HST CANDELS \citep{CANDELS1,CANDELS2} does overlap with the JADES DR2 GOODS-S imaging and would likely improve photometric redshift performance by extending wavelength coverage to bluer bands and leveraging spectral breaks not seen in the NIRCam filters.  \textsc{DeepDISC} requires a common pixel grid for all input imaging, so the inclusion of CANDELS data would be limited to HST ACS imaging (rather than WFC3). However, our goal is to isolate and quantify the performance of \textsc{DeepDISC} using JWST imaging alone, motivated by forthcoming JWST programs that will not necessarily have homogeneous ancillary HST or ground-based coverage. For this reason, we deliberately restricted our fiducial analysis to the JADES JWST imaging to ensure a uniform and self-contained dataset.  However, joint-instrument photo-z inference and incorporation of data without common exploration into multimodal  will be followed up in future work (e.g., in the  COSMOS field).

We compile spectroscopic redshifts from the third JADES data release \citep[JADES DR3;][]{JADESDR3}, ASTRODEEP-GS43 catalog \citep{ASTRODEEP}, and the second data release of MUSE Hubble Ultra Deep Field surveys \citep[MUSE-HUDFS;][]{MUSE-HUDFS}. Redshifts reported in JADES DR3 were measured from the combined NIRSpec low-resolution CLEAR/PRISM and medium-resolution F070LP/G140M, F170LP/G235M and F290LP/G395M observations \citep{JADESDR3}. We keep redshifts with no \texttt{DR\_flag=False} and \texttt{z\_Spec\_flag=A or B or C} to ensure secure redshift. \cite{ASTRODEEP} compiled a list of 4951 high quality and publicly available spectroscopic redshifts in the CANDELS GOODS-South field. For spectroscopic redshifts from MUSE-HUDFS, we reject sources with low redshift confidence (\texttt{zconf=1}) and low matching confidence (\texttt{Mconf=1}). We associate spectroscopic redshifts with photometric objects from JADES DR2 and adopt a 0\farcs5 crossmatch radius for MUSE-HUDFS and ASTRODEEP-GS43 catalog. For spectroscopic redshifts with multiple photometric sources within the matching radius, we determine the association by choosing the brightest object in the HST F775W filter for MUSE-HUDFS and F160W for the ASTRODEEP-GS43 catalog. After excluding 12 objects with significant redshift difference ($\Delta z/(1+z)>0.1$) between the MUSE-HUDFS and the ASTRODEEP-GS43 catalog, we obtain reliable spectroscopic redshifts for 4341 objects.

\begin{figure*}
    \centering
    \includegraphics[width=\textwidth]{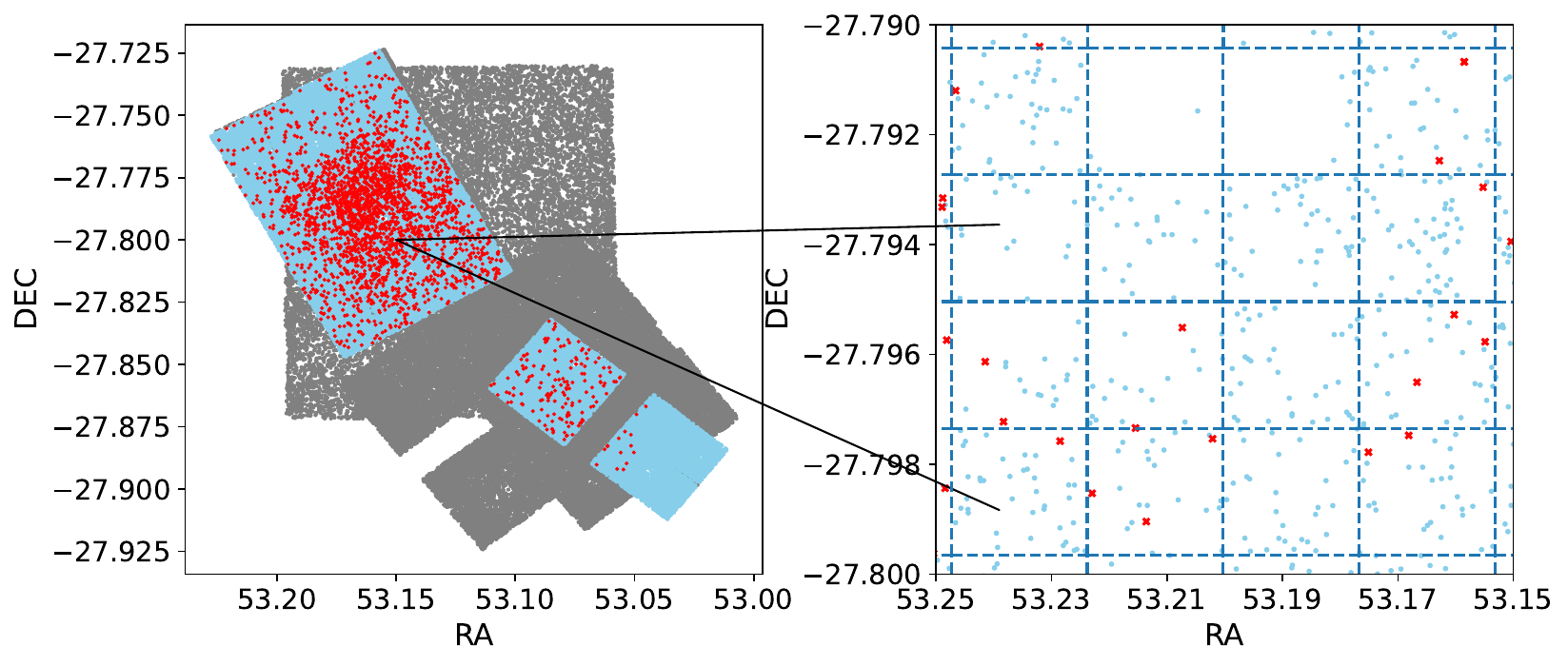}
    \caption{JADES photometric and spectroscopic catalog footprint. The left panel shows all sources in the photometric catalog (grey) and sources where imaging is present in the 9 NIRCam filters from DR1 (blue). Red points indicate our spectroscopic sample.  The right panel shows how the imaging data is partitioned into sub-images that can contain multiple sources with spec-zs.  We exclude from our training set any sub-image that contains no sources with spec-zs.}
    \label{fig:footprint}
\end{figure*}

\begin{figure}
    \centering
    \includegraphics[width=0.5\textwidth]{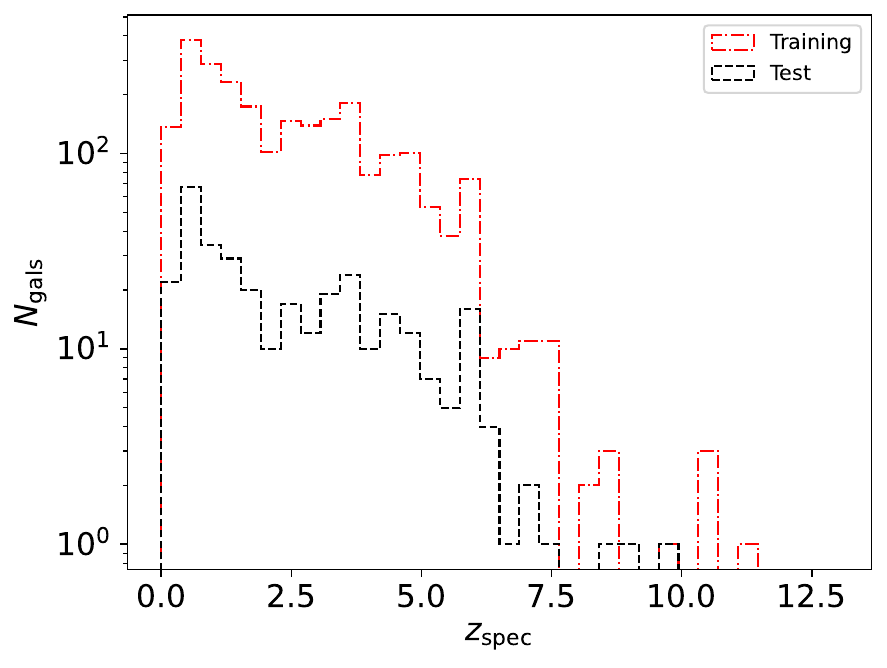}
    \caption{N(z) for objects in the training and test sets. Few sources exist at $z>6$ in either sample.}
    \label{fig:nz}
\end{figure}

To derive our datasets for training and testing, we partition the imaging data into sub-images of equal size.  Each sub-image is 8.3 by 7.5 arcseconds.  To train the network, we require each image to have corresponding ground truth information composed of object locations, segmentation maps, bounding boxes, and redshifts.  Object locations, segmentation maps and bounding boxes are taken from the segmentation image and photometric catalog.  Due to the fact that the JADES imaging is created from composite observations from multiple programs, the sky coverage of each filter is not completely overlapping. This is particularly evident for F335M, where program 1180 does not extend the coverage of program 1210. We exclude any sub-image that does not have sky coverage in the 9 filters listed above.  This is done so that the network maximizes the use of all data and does not potentially confuse a lack of imaging data with a dropout in a given filter.  We are left with  2145 objects with a reliable spec-z and photometry in the 9 NIRCam filters. The footprint of the photometric catalog, as well as the footprint of spectroscopic objects with complete 9-filter coverage is shown in Figure \ref{fig:footprint}.  We also exclude any sub-image without at least one object therein with a spectroscopic redshift.  Once each sub-image has its corresponding ground truth data, we divide the dataset into training, test, and validation sets. The redshift distribution of our training and test sets is shown in Figure \ref{fig:nz}.  We randomly partition 90\% of the images into a training set and 10\% of the images into a test set.  A further 10\% of the training set is held out for validation.  The training set contains 1845 objects with a spec-z, and the test set contains 330.

One may see two potential pitfalls of using a deep learning algorithm given the data.  First, redshift coverage drops off sharply at $z>6$, with the highest redshift source at $z\sim 11$.  Previous development of photo-z estimators has shown machine learning methods tend to under-perform in underrepresented regimes of the training sample.  Second, our dataset is in general fairly small for training a neural network, which leaves our method prone to overfitting, or memorizing the training data at the expense of poor performance on the test set. We discuss steps taken to prevent overfitting and the implications of using \textsc{DeepDISC} photo-zs on larger samples in later Sections.

\section{Methods}
\label{sec:methods}
Our photometric redshift estimator is built on the \textsc{\textsc{DeepDISC}} instance segmentation framework; we summarize its operation and our extensions here. \textsc{\textsc{DeepDISC}} is based on the framework outlined in \cite{Merz2023} and \cite{Merz25}.  It is an instance segmentation model, designed for astronomical images.  Instance segmentation consists of detecting objects, classifying them and deblending by producing a segmentation mask for each object.  Instance segmentation models have been employed in many different contexts, from remote sensing \citep{Gui24} to medical imaging, to self driving cars and more.  \textsc{DeepDISC} is a two-stage detection model, consisting of a backbone network, a Region Proposal Network (RPN), and Region of Interest (ROI) heads. The backbone network extracts features from input images. The RPN uses these features to learn where objects are in the images, based on a supplied ground truth.  The proposed regions from the RPN are used to select features from the backbone that correspond to the proposed objects.  Proposed objects are then matched to ground truth objects using an intersection-over-union (IOU) criterion.  IOU is calculated by taking the intersectional area of a proposed object's bounding box and a ground truth object's bounding box and dividing by the union area.  If this quantity meets a predefined threshold, the object is considered detected and the corresponding features are sent to the ROI heads, which perform downstream measurements such as classification and segmentation.

\cite{Burke_deblending_2019} first used a Mask-RCNN instance segmentation model on astronomical images, demonstrating the usefulness and applicability of this framework on large-scale images from astronomical surveys.  \cite{Merz2023} developed \textsc{\textsc{DeepDISC}}, an extension of this framework interfaced with \texttt{detectron2}, a repository of instance segmentation models that includes state-of-the-art methods.  In that work, they conduct a comparison study of different models and different image processing techniques, using Hyper Suprime-Cam Subaru Strategic Program (HSC SSP) images from Data Release 3 \citep{aihara_third_2022} to benchmark detection, deblending, and star/galaxy classification.  

The \textsc{\textsc{DeepDISC}} framework is further expanded in \cite{Merz25} to include photometric redshift estimation and included within the Redshift Assessment Infrastructure Layers package \citep{RAIL25} for the LSST Dark Energy Science Collaboration.  By adding a Mixture Density Network (MDN) to the model, photo-zs are parametrized with a Gaussian mixture model.  A probability density function (PDF) is produced for each detected object, allowing us to characterize uncertainty in our redshift measurement.  Milky Way dust extinction is included as an extra neuron input to the MDN, by using the \texttt{dustmaps} \citep{Dustmaps} package and image world coordinate system to get the E(B-V) value at the locations of the objects in the training set.  The redshift ROI head is trained to minimize the negative log likelihood of the training redshifts given the proposed MDN.  We adopt the same redshift ROI head as in \cite{Merz25}.  The total loss function of the network includes the loss functions of the other ROI heads and the loss from the RPN. It is written as 
\begin{equation}
    L_{\rm total} = L^\alpha_{\rm box} + L^\beta_{\rm box} + L^\beta_{\rm class} + L^\beta_{\rm mask} + L^\beta_{\rm redshift}
    \label{eq:loss}
\end{equation}
where the superscript $\alpha$ indicates a term calculated from the RPN output, and a superscript $\beta$ indicates a term calculated from the ROI head outputs.  $L_{\rm box}$ is calculated as the mean-squared error of predicted bounding box corners vs ground truth bounding box corners,  $L_{\rm class}$ is the categorical cross entropy loss between predicted and true classes, $L_{\rm mask}$ is determined from the average binary cross-entropy loss of the predicted/ground truth masks, and $L_{\rm redshift}$ is the negative log likelihood. We use a single class, "object" for our labels, which was shown in \cite{Merz25} to have little effect on predicted photometric redshifts.  We train for 150 epochs, and use early stopping to avoid overfitting.  The training optimizer is the stochastic gradient descent method, with an initial learning rate of 1e-3.  The learning rate is reduced by a factor of 10 after 45, 75, and 105 epochs.  We normalize all input images by a z-score, i.e., we divide each image by the pixel mean of the training sample and divide by the pixel standard deviation.  

After training, the network predicts a bounding box, segmentation mask, class and redshift for each object.  Objects are matched to the JADES DR2 GOODS-S catalog by calculating the IOU of the \textsc{DeepDISC} predicted boxes with the DR2 bounding boxes produced from the segmentation image.  The IOU threshold is set to 0.3 to aid in small object detection. 

\subsection{Model Backbones and pretraining}

The backbone feature extractor is an important part of instance segmentation models and the \textsc{\textsc{DeepDISC}} framework.  A variety of neural networks can be used as a backbone, including convolutional neural networks such as ResNets \citep{He2016}.  ResNets are a common deep neural network architecture used in a variety of domains.  They utilize a convolutional operation to learn spatial inductive bias, and utilize skip connections to allow deeper layers to more easily learn from earlier layers. ResNets have been demonstrated to work with low-redshift galaxy photo-z estimation in \cite{Hayat21}.  

A new paradigm of deep learning was introduced by \cite{Attention} with the transformer.  Rather than convolutional operations, transformers utilize an attention mechanism by encoding an input tensor into key, query, and value tensors, on which combination of linear and nonlinear operations are performed.  Keys, queries and values thus capture relationships across the entire input tensor.  This architecture is the foundation of modern large language models like Chat-GPT and has been recently adopted for computer vision 
\cite{Dosovitskiy20}.  \cite{Merz2023} tested \textsc{\textsc{DeepDISC}} classification, detection, and deblending with both ResNet and state-of-the-art vision transformer backbones.

All models in \cite{Merz2023} are pretrained on terrestrial images from the ImageNet1k dataset \cite{ImageNet}.  Under these test conditions, models based on transformers \citep{mvitv2,Liu21} are found to be more robust to image processing techniques and generally outperformed ResNets in star/galaxy classification, object detection and segmentation.  \cite{Merz25} use a Multi-scale Vision Transformer \citep[MViTv2][]{mvitv2} to produce \textsc{DeepDISC} photo-zs on simulated LSST data.  The MViTv2 utilizes pooling and hybrid window attention, operations designed to capture global and local information, as well as reduce computing complexity.

The MViTv2 backbone used with \textsc{\textsc{DeepDISC} photo-z} in \cite{Merz25} performs on par with or better than traditional photo-z estimation methods on a large sample of objects from LSST DESC simulations \citep{DC2}.  However, in this work, we continue to test the effectiveness of ResNet backbones in the \textsc{\textsc{DeepDISC}} framework, for a few reasons.  Vision transformers, while powerful, are computationally expensive and require very large training sets to fully realize their potential, whereas ResNets train faster and are less prone to overfitting on small datasets \citep{Dosovitskiy20}.  Transformer models are inherently larger than CNNs, and are known to be ``data hungry" and typically require a large amount of data to achieve optimal performance.  While the JWST datasets used in this work are rich in information due to their high resolution, broad wavelength coverage, and depth, we are limited by the number of available spectroscopic redshifts for supervised training.  Even with a large number (O(10$^5$)) of redshifts available for training with simulated LSST data, \cite{Merz25} found that \textsc{\textsc{DeepDISC}} with a transformer backbone was sensitive to the relative lack of redshifts beyond z$\sim$2.5 in the data.

While vision transformers proved viable for photo-z estimation on large simulated datasets, the backbones employed in \cite{Merz2023} and \cite{Merz25} are pretrained on terrestrial images.  The detectron2 library on which \textsc{DeepDISC} is built provides several open-source models that are pretrained on ImageNet and MS-COCO datasets.  These models have been trained to classify everyday objects such as animals, cars, etc., in RGB images.  \cite{Merz2023,Merz25}, and the initial work of \cite{Burke_deblending_2019} utilize transfer learning to let the networks use knowledge learned from terrestrial images as a ``starting point" to then learn from astronomical images. However, pretraining using images of galaxies may improve our model performance.  Supervised pretraining on galaxy images has been shown to increase performance in classifying galaxy types and morphologies compared to pretraining with ImageNet data \cite{Walmsley22,Walmsley24}. \cite{Hayat21} show that pretraining a ResNet on galaxy images with contrastive learning boosts downstream photo-z estimation.  Pretraining with galaxy images instead of ImageNet has also improved detection of giant star-forming clumps with a Faster R-CNN model \cite{Popp24}. It is possible that pretraining on images outside of the astronomical domain introduces a bias that convolutional neural networks have a difficult time mitigating, considering the inherent differences in noise properties and dynamic ranges of astronomical images vs terrestrial RGB images.  This effect is more pronounced in our case, as we use the multiband images as input for photo-z estimation, rather than converting to RGB, as is frequently done in studies of galaxy morphology \citep{Walmsley24}. Transformers may be more capable of overcoming this difficulty due to their global attention mechanism, which allows them to learn features corresponding to large scales, and weaker inductive biases. However, this learning is limited in small data regimes, where CNNs may thrive due to their stronger inductive biases \citep{Dosovitskiy20}.  For a targeted application of \textsc{DeepDISC} on a small dataset of JWST images, we test this trade-off between a model that can learn generalizable features from sufficient amounts of data (MViTv2 backbone) and a model that can use a stronger prior on the data structure to perform well with limited data (ResNet backbone).

\section{Results}\label{sec:results}

\begin{figure*}[htb!]
    \centering
    \subfloat{\includegraphics[width=0.48\textwidth,height=7cm]{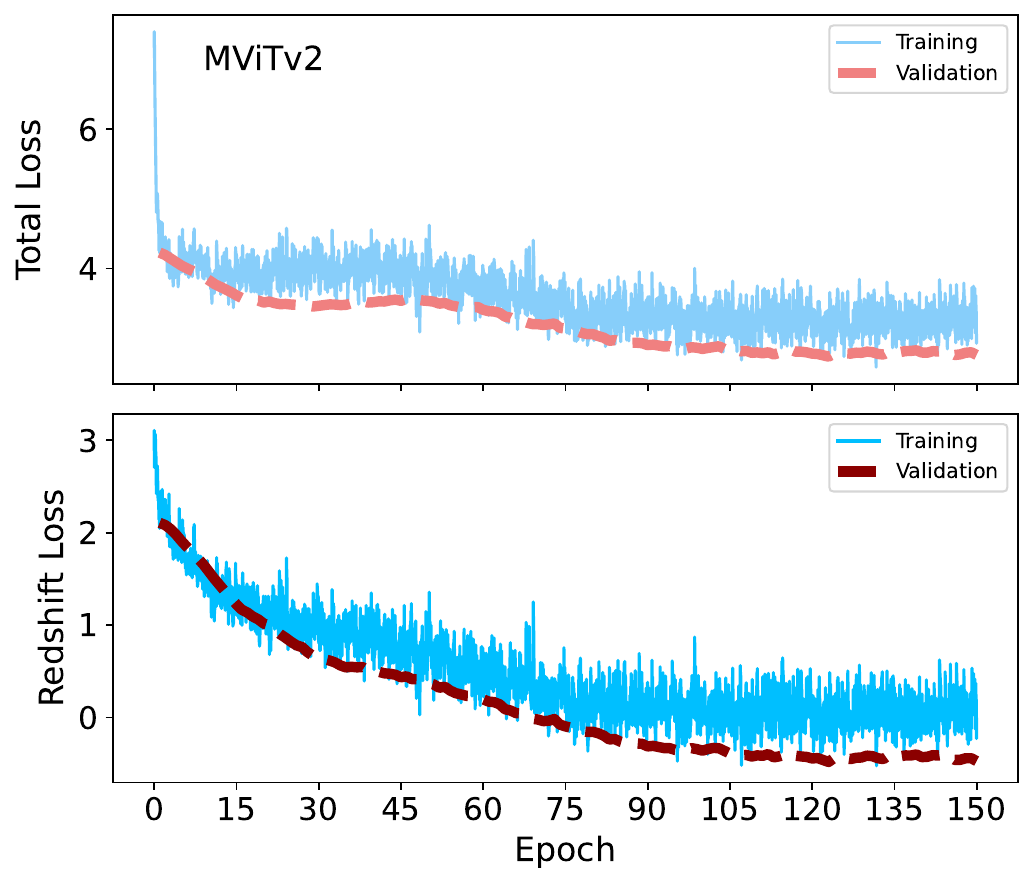}}
    \quad
    \subfloat{\includegraphics[width=0.48\textwidth,height=7cm]{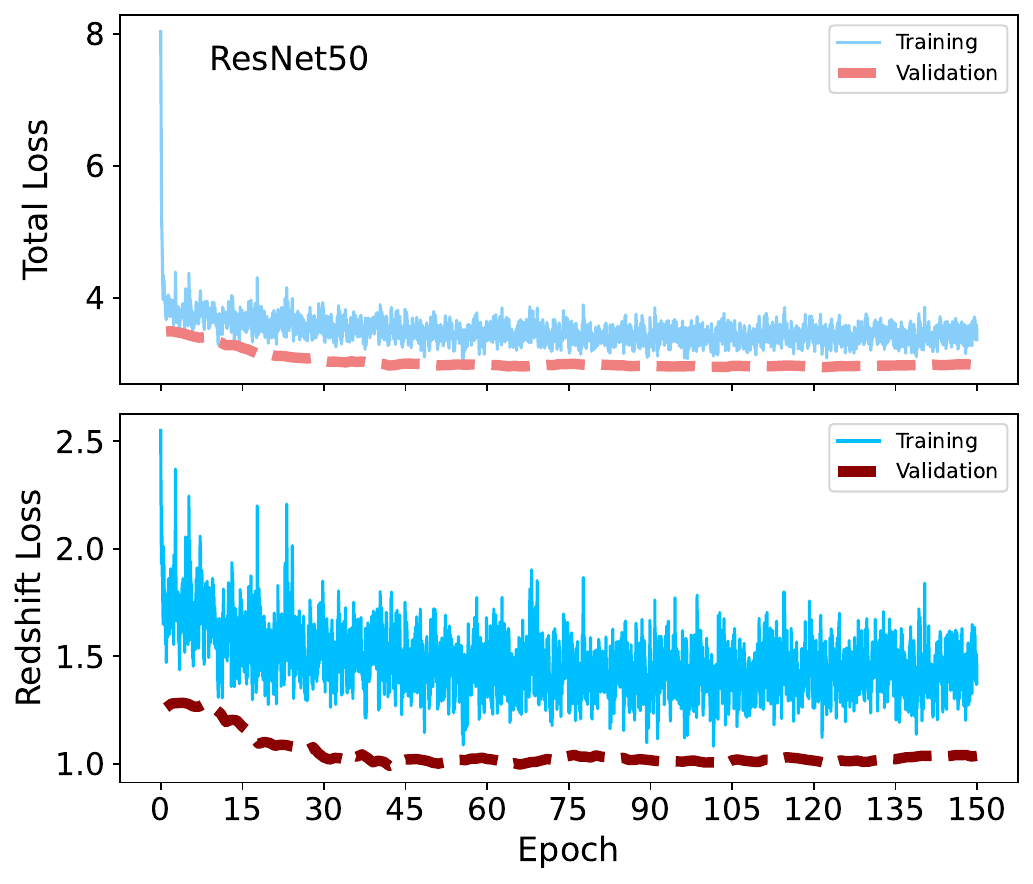}}
    \caption{Loss curves showing the training and validation set loss for two models pretrained on ImageNet data over 150 epochs. The MViTv2 network (left) consistently learns over the course of the training, as the training and validation loss decrease over time.  The ResNet50 does not learn as much with regards to object redshift, evidenced by the higher validation set loss.  The losses have been smoothed with an exponential moving average.  The upper panels show the total loss, defined in Equation \ref{eq:loss} and the lower panels show only the redshift component of the loss.}
    \label{fig:IMpre_loss_curves}

\end{figure*}

\begin{figure*}[htb!]
    \centering
    \subfloat{\includegraphics[width=0.48\textwidth,height=7cm]{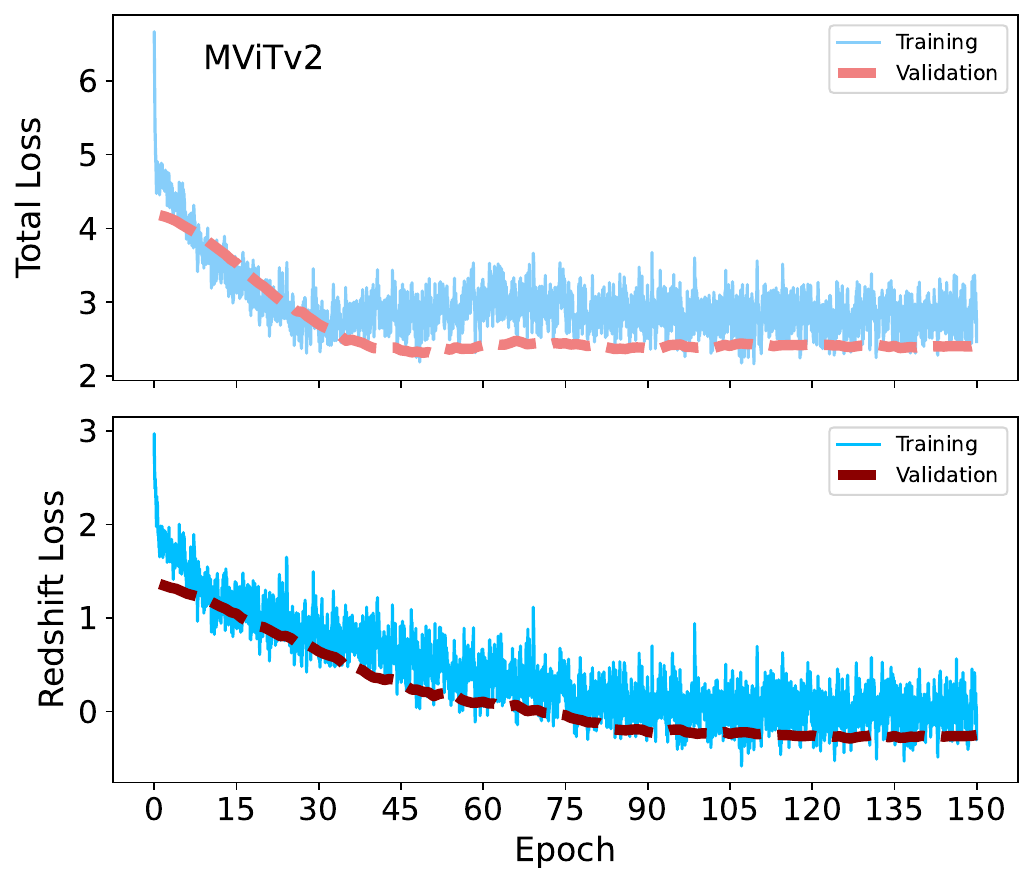}}
    \subfloat{\includegraphics[width=0.48\textwidth,height=7cm]{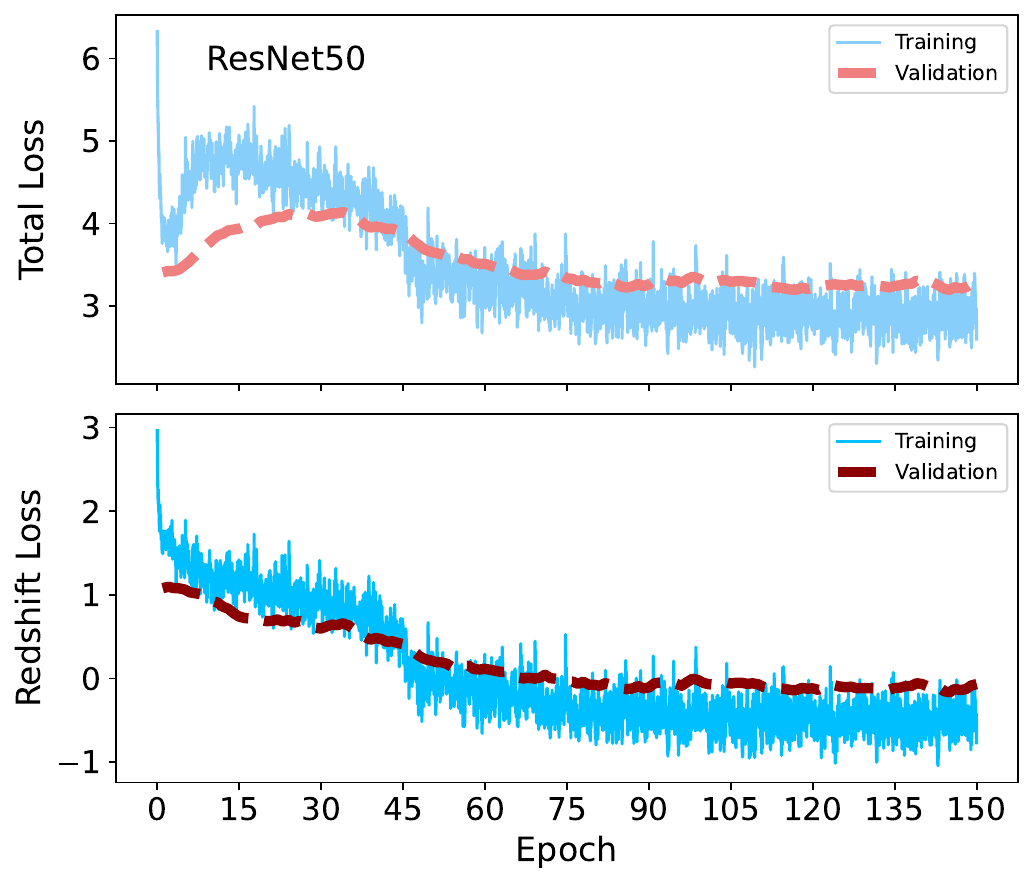}}
    \caption{Loss curves showing the training and validation set loss for for two models pretrained on Galaxies ML data over 150 epochs. The MViTv2 network (left) curves appear qualitatively similar to those in Figure \ref{fig:IMpre_loss_curves}, indicating that pretraining does not have a large effect.  The ResNet50 network (right) achieves lower validation set redshift loss compared to the previous result. Loss curves are smoothed and arranged as in Figure \ref{fig:IMpre_loss_curves}.}
    \label{fig:GMLpre_loss_curves}

\end{figure*}

\begin{figure*}[htb!]
    \centering
    \includegraphics[width=0.66\textwidth]{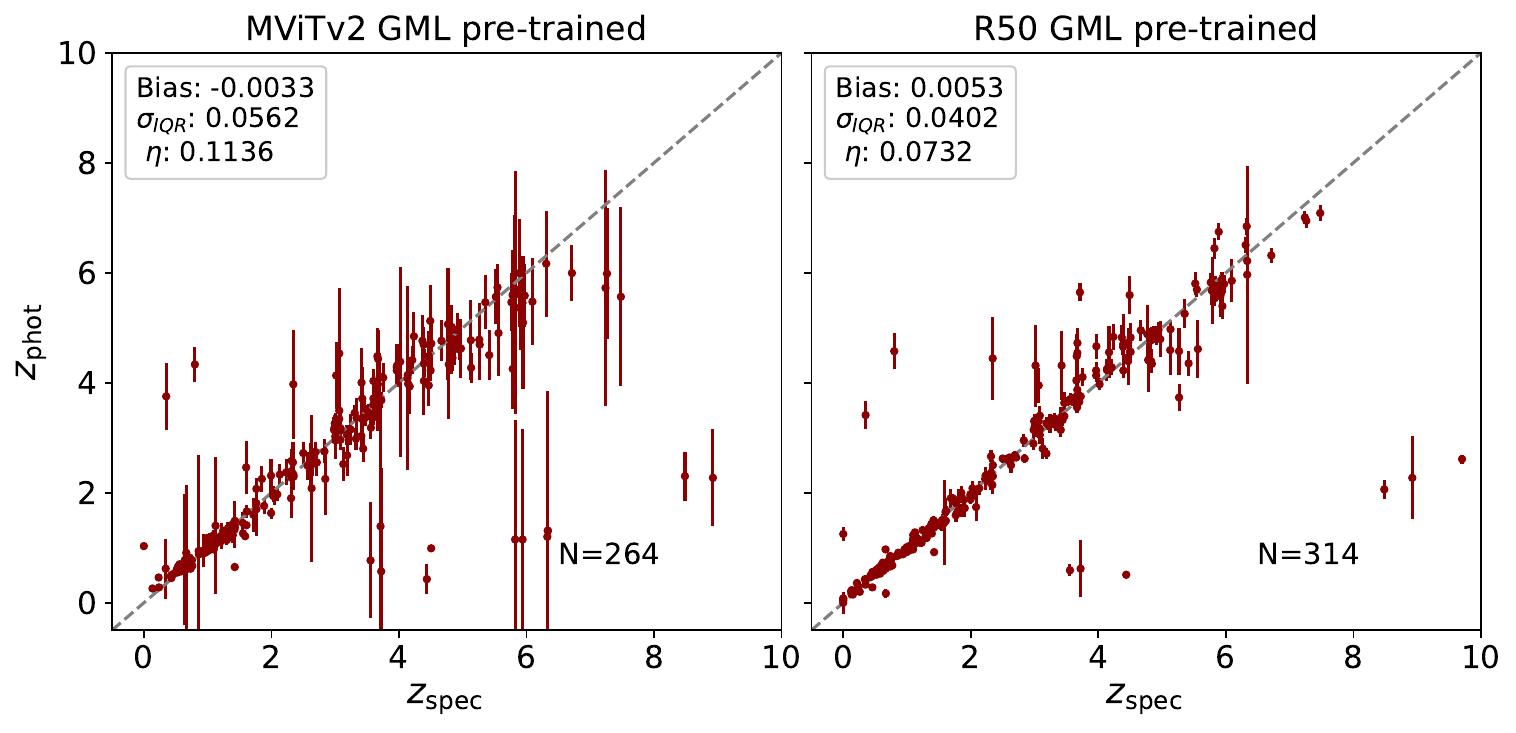}
    \caption{Photo-z point estimate scatter plot comparing an MViT and ResNet50 (R50) model. Both models are pretrained using the GalaxiesML dataset of images and redshifts from HSC DR2. The mode of each PDF is used as a point estimate $z_{\rm phot}$ and error bars are taken from the 68\% confidence intervals.  The number of detected objects is indicated in the bottom right. The R50 model outperforms the MViTv2 model, in terms of bias, scatter $\sigma_{\rm IQR}$ and outlier fraction $\eta$ as well as object detection.}
    \label{fig:model_comp_scatter}
\end{figure*}

\begin{figure*}
    \centering
    \includegraphics[width=\textwidth]{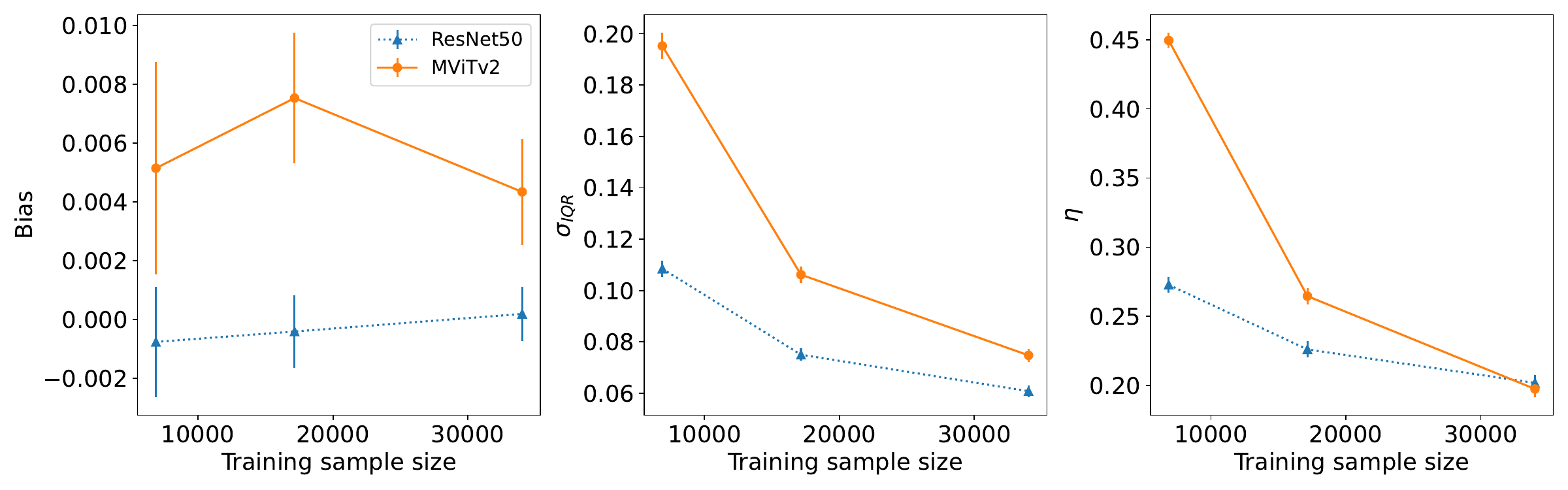}
    \caption{Point estimate metrics as a function of training data size for JAGUAR simulation data.  Errors are taken from bootstrapping the test set.  While both the ResNet50 and MViTv2 model scale in performance with increasing dataset size, the ResNet50 model is much less sensitive to smaller dataset sizes.}
    \label{fig:jag_scaling_metrics}
\end{figure*}

\begin{figure}
    \centering
    \subfloat{\includegraphics[width=\columnwidth]{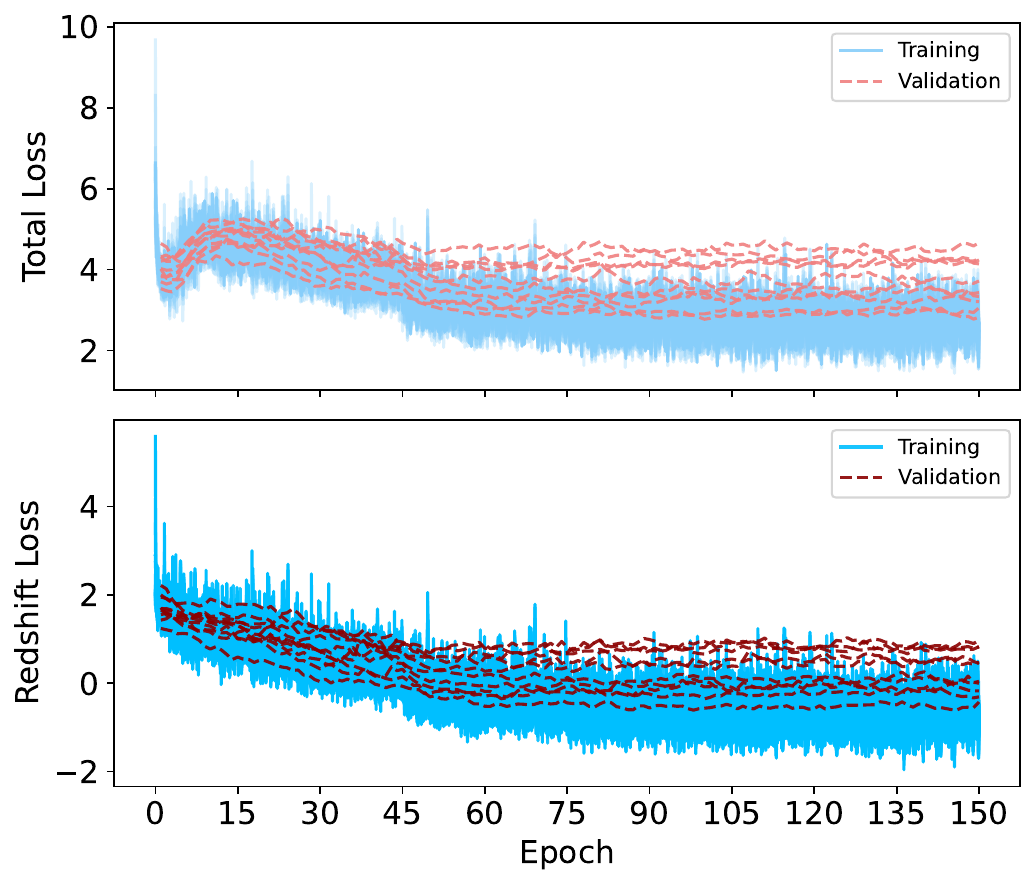}}
    \quad
    \subfloat{\includegraphics[width=\columnwidth]{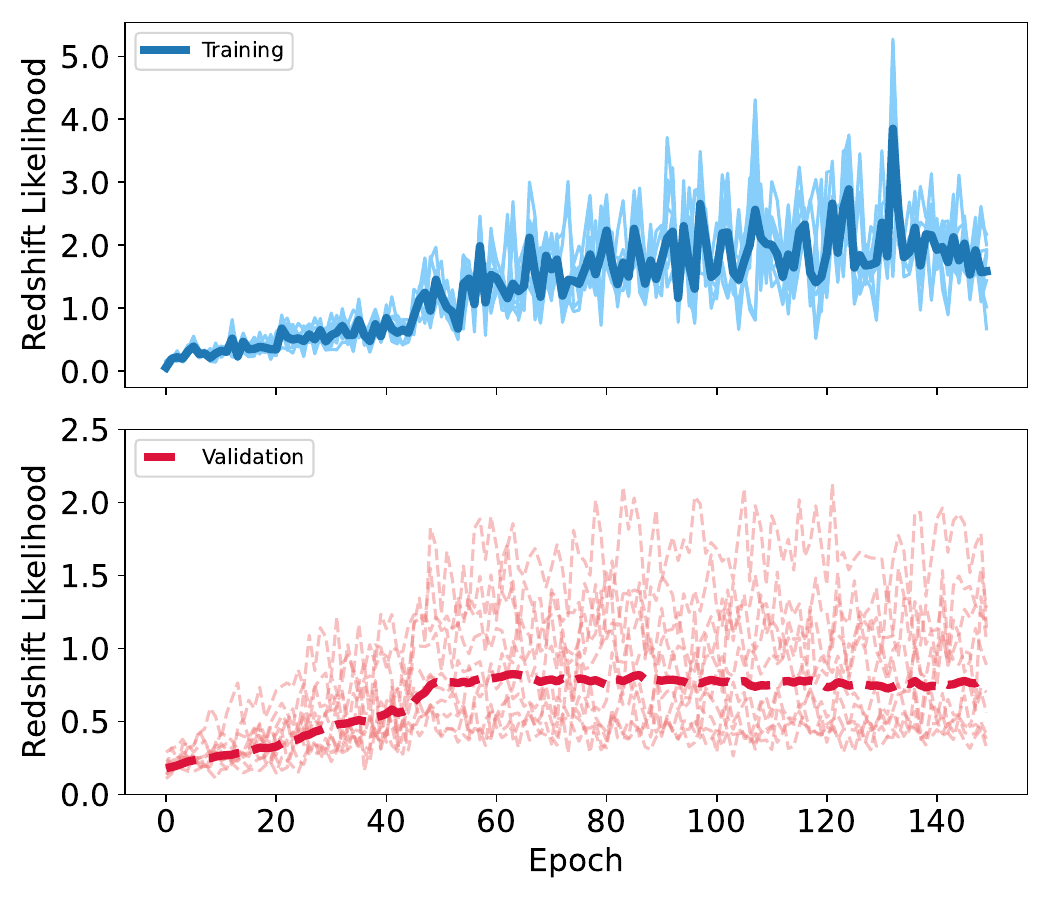}}
    \caption{Upper panels: training and validation loss for an ensemble of models, using the GML-pretrained ResNet50.  We use the same visualization as in Figures \ref{fig:IMpre_loss_curves} and \ref{fig:GMLpre_loss_curves}. Overfitting appears around epoch 70.  Lower panels: training and validation redshift likelihood for an ensemble of models, using the GML-pretrained ResNet50. The averages across the ensemble are shown as the bold, thicker lines. }
    \label{fig:ensloss}
\end{figure}

\begin{figure*}
    \centering
    \includegraphics[width=\textwidth]{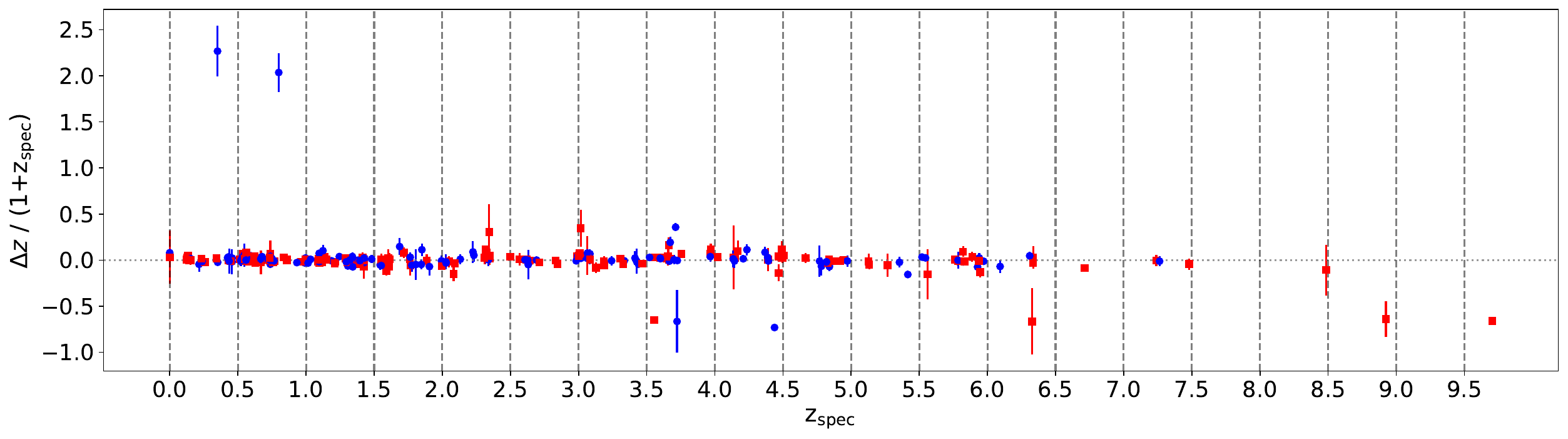}
    \caption{Weighted scatter in $z_{\rm spec}$ vs ($z_{\rm phot}$-$z_{\rm spec}$)/(1+$z_{\rm spec}$).  Markers correspond to object magnitudes, and are relative within each bin, shown as the dotted grey lines. Red squares indicate F444W magnitudes fainter than or equal to the bin median, and blue circles indicate F444W magnitudes brighter than the bin median.}
    \label{fig:scatter_bins}
\end{figure*}

\begin{figure*}
    \centering
    \includegraphics[width=\textwidth]{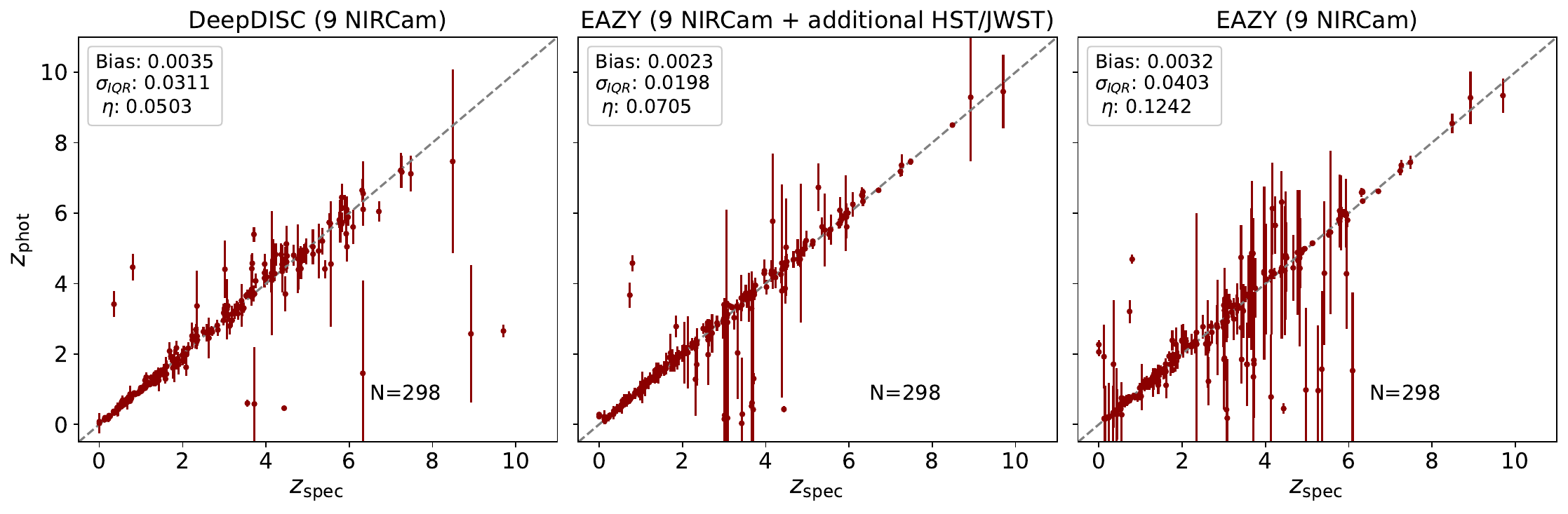}
    \caption{Left: $z_{\rm spec}$ vs $z_{\rm phot}$ for objects detected by our \textsc{DeepDISC} ensemble. \textsc{DeepDISC} $z_{\rm phot}$ is taken as the mode of the PDFs and error bars are derived from the 68\% confidence intervals.  We list the number of detected objects in our test set (298 out of 330).  Middle: \texttt{EAZY} photo-z estimates from the JADES DR2 GOODS-S catalog for the same set of objects. The JADES DR2 catalog utilizes additional JWST imaging from JEMS and HST ACS where available.  For \texttt{EAZY}, we use the redshift at minimum chi-square \texttt{EAZY}\_z\_a along with upper and lower 68\% confidence intervals for error bars. Right: \texttt{EAZY} estimates using the photometry from the JADES DR2 catalog, but limited to the 9 filters seen by \textsc{DeepDISC}. Compared to \texttt{EAZY} estimates that use more color information, \textsc{DeepDISC} yields a larger bias and scatter, but smaller outlier fraction.  If the color information between the two algorithms is the same, \textsc{DeepDISC} yields better point estimates.  \texttt{EAZY} produces more accurate estimates at z>8, although small number statistics likely dominate results in this regime.}
    \label{fig:EZ_comp_scatter}
\end{figure*}

\begin{figure}[htb!]
    \centering
    \includegraphics[width=0.9\columnwidth,height=21cm]{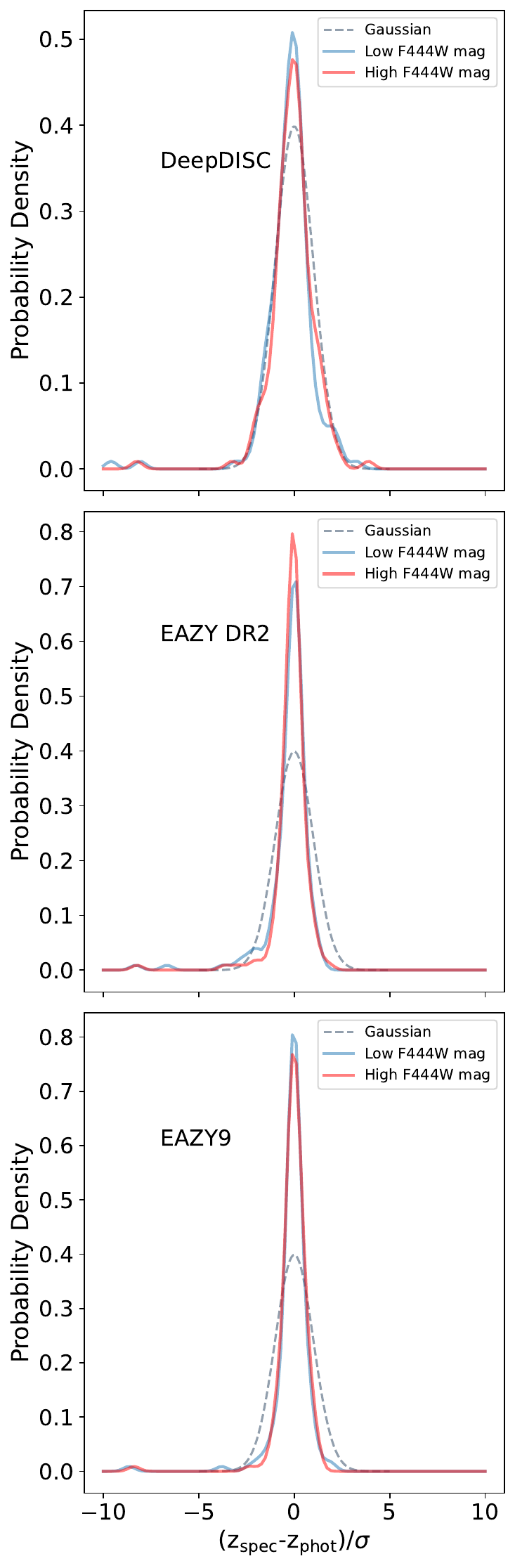}
    \caption{Kernel density estimation of the histogram of $z_{\rm phot}$-$z_{\rm spec}$, weighted by the error estimate (68\% confidence interval) from the PDFs.  The histograms are produced from the 298 objects in the test set commonly detected by all models in the ensemble. The sample is split magnitudes less than or greater than the median F444W Kron magnitude. A normal distribution is shown for reference (grey dashed line).  \textsc{DeepDISC} results are shown in the top panel, \texttt{EAZY} results from the JADES DR2 catalog are shown in the middle, and \texttt{EAZY} with 9 NIRCam filters are shown on the bottom.}
    \label{fig:Norm_error_hists}

\end{figure}

\begin{figure}
    \centering
    \includegraphics[width=\columnwidth]{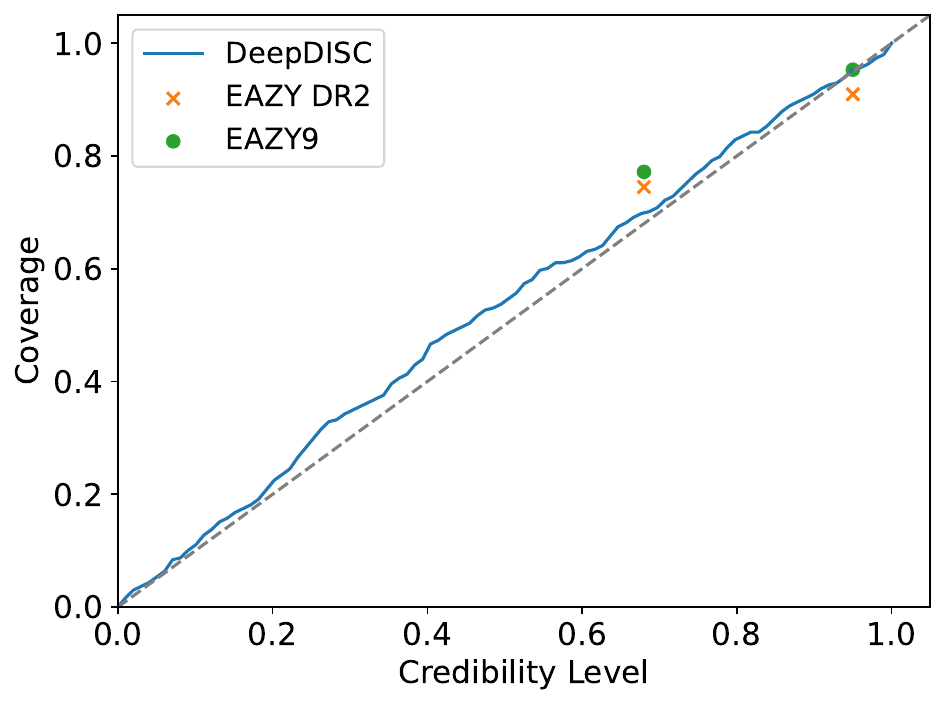}
    \caption{Coverage of \textsc{DeepDISC} photo-z posterior PDFs vs credibility level. The blue line ideally follows the diagonal, where the fraction of PDFs that capture the $z_{\rm spec}$ given a credibility level is equal to that credibility level.  \texttt{EAZY} credibility levels are shown as points.}
    \label{fig:coverage_intervals}
\end{figure}

We first train two \textsc{\textsc{DeepDISC}} models on the JADES training set using ImageNet pretrained weights to establish a baseline. One model uses a MViTv2 backbone, and another uses a ResNet50 backbone. The MViTv2 is a newer iteration of the Multiscale Vision Transformer network, and a ResNet50 is a ResNet architecture with 50 total layers.  Loss curves for these models are shown in Figure \ref{fig:IMpre_loss_curves}.  As expected, the MViTv2 model is able to adapt to the new domain of JWST images.  Training and validation set loss decrease over time (number of epochs), indicating that the model is actively learning.  Conversely, the ResNet model does not adapt well to the new domain, as losses are generally stagnant and the redshift component of the loss remains high.  This is consistent with previous results from \citep{Merz2023}, which show that ResNets pretrained on ImageNet tend to perform poorly on astronomical data when tasked with detection, deblending, and classifying stars/galaxies, unless an asinh pixel scaling is used.  \cite{Merz25} have also shown that an out-of-the-box MViTv2 backbone leads to photo-z estimation that surpasses traditional catalog photometry-driven methods.  It appears likely that inductive biases learned during training with terrestrial images are too strong for the ResNet backbones to overcome when adapting to astronomical images.  However, in-domain pretraining provides an avenue for overcoming this limitation. 

\subsection{Pretraining and Architecture Selection}
Pretraining on galaxy images lets the network learn features general to the target domain, and creates a prior for the network at the start of the training phase.  This process can be computationally expensive, but it is worth the effort.  \cite{Walmsley24} show that in-domain pretraining is an important step in downstream tasks related to galaxy morphological classification; models that are pretrained on galaxy images outperformed those trained from scratch or pretrained on ImageNet alone.  \cite{Erikson20} estimate photo-zs for the Physics of the Accelerating Universe Survey data (albeit with aperture fluxes instead of images as input) and find improved performance when pretraining with simulated data compared to no pretraining.  Here, we test these insights in the context of image-based photo-z estimation.  

For in-domain pretraining, we use the Galaxies ML dataset (hereafter GML) from \cite{Do2024}.  This dataset consists of HSC grizy images of 204,573 galaxies with spectroscopic redshifts. It is arguable that this dataset is not truly ''in domain" with the JWST data, as it consists of ground-based optical imaging instead of space-based NIR imaging.  Colors, point spread functions, resolutions, noise, etc. will be fundamentally different between these datasets.  However, we pick this dataset because it contains carefully curated images with corresponding spec-zs.  For supervised pretraining, reliable ground truths are necessary to ensure an accurate mapping of input pixels to redshift is learned, and we do not have access to a large corpus of real images in the JADES filters with corresponding spectroscopic redshifts.  For general applications to a given astronomical survey, it is not necessarily practical to obtain large pretraining datasets of images from the same survey.  The need for so-called ''foundation models" that can be applied to many different use-cases and datasets is quickly becoming more apparent in astronomy research \citep{Walmsley24, AstroCLIP,Euclid-AstroPT}.  Additionally, while pretraining has been shown to improve knowledge transfer through feature reuse, pretrained weights can be used to capture ``low-level" statistical properties of images in the target domain that contain useful information \citep{Neyshabur20}. Thus, we pick the GML dataset for its usefulness in pretraining for photo-z estimation and letting the networks establish a prior such that initial features are inherently correlated with redshift.  Further investigation into using backbones pretrained with different astronomical imaging datasets, and/or on different tasks (supervised or unsupervised) is worth exploring, but outside the scope of this work.  We pretrain the MViTv2 and ResNet50 backbones to produce point-estimates for each galaxy's redshift.  During pretraining, we attach a multi-layer perceptron consisting of two layers to each backbone, which will take the features from the deepest layers of the backbone networks, and output a single value for the predicted redshift.  We pretrain these backbones with a stochastic gradient descent optimizer and mean squared error loss for 50 epochs.  

We note that due to the difference in available filters (i.e., image color channels) for the JADES and GML or ImagNet datasets, the first layer of the pretrained models will be re-initialized with random weights during training.  This was also the case for the model used in \cite{Merz25} and our ImageNet test cases above.  Given that the early layers of the MViTv2 transformer backbone are able to capture larger-scale correlations \cite{Dosovitskiy20}, the ability to quickly adapt to the new color channels is not surprising.  It is unclear, however, if a ResNet50 backbone will also be able to adapt to this difference in data, especially given the results from ImageNet pretraining.

After GML pretraining is complete, we use the new backbone weights to initialize the models for training with the JADES data. Training loss curves for the GML pretrained models are shown in Figure \ref{fig:GMLpre_loss_curves}.  We again see the MViTv2 model learn over time, with the validation loss plateauing to roughly the same value as the ImageNet pretrained model, indicating that performance is roughly independent of the pretraining dataset.  This means that inductive biases learned by the MViTv2 model during pretraining are indeed relatively weak.  In contrast, the ResNet50 model now performs much better, as the validation loss plateaus to a lower value.  The inductive biases inherent to the convolutional nature of the ResNet, e.g., spatial locations and translational invariance, likely help transfer knowledge from galaxy images in one survey to another.  While galaxies observed at different resolutions, wavelengths and redshifts will appear very differently, common high-level properties such as bulges/disks and the general relationship between color and redshift were likely learned during the pretraining phase.  We determine that the GML-pretrained R50 model yields more accurate photo-zs than MViTv2 models pretrained on either GML or ImageNet. We also test the case of no pretraining (i.e., using random weight initialization), which leads to roughly the same results with MViTv2 and worse results with the R50 model (see Appendix \ref{app:MI}).  This again evidences the generalizability of transformer models, but their poor specialization in small data regimes.

To further demonstrate the sensitivity of the models to the size of the training data, we train the GML pretrained models with simulated data from the JADES extraGalactic Ultradeep Artificial Realizations \citep[JAGUAR][]{JAGUAR}.  The JAGUAR model uses empirical stellar mass and UV luminosity functions measured from $0<z<10$.  Stellar mass functions are modeled to generate populations of quiescent and star-forming galaxies, and spectra modeled with \textsc{BEAGLE} \citep{BEAGLE} are assigned based on empirical relations to properties such as UV absolute magnitude and stellar mass.  Morphologies are assigned as a single Sersic profile, with half light radii, axis ratios and Sersic indices chosen based on empirical relations to UV luminosity, stellar mass, and redshift.  Images are created using the \textit{Guitarra} image simulation tool and contain systematics such as zodiacal and telescope background light, cosmic rays, read noise, and the detector signatures as measured from ground-based data. Scenes are created from dither positions calculated for JWST program 1180, corresponding to the first JADES data release, and are combined to match the depth of the JADES survey. 

We divide the JAGUAR images into sub-images of approximately the same size as our JADES data.  Within each sub-image, we produce ground truth bounding box and segmentation mask annotations by individually simulating every object in the sub-image, and thresholding based on a noise level.  Objects morphologies are modeled with a Sersic profile, which we draw using GalSim.  We draw the profiles in each JADES filter using the fluxes and shape parameters listed in the JAGUAR catalog, and convolve with the PSFs used for the simulated images, which were created with the WebbPSF software.  The noiseless, convolved profiles are then thresholded at 2$\sigma$ above an empirical background obtained from sigma-clipping the images.  

Since we are limited in our sample of real JWST data, we use the simulations to extrapolate how the models will behave with larger datasets.  We divide the JAGUAR image data into three training samples of differing size, and one test set.  We use the GML pretrained backbones for our test.  After a model is trained on each of the three training sets, we compute bias, scatter, and outlier fractions of photo-zs for detected objects in the test set.  We take the common set of detected objects for all three models to remove any sample selection biases.  Results are shown in Figure \ref{fig:jag_scaling_metrics}.  In general, the ResNet50 model performs better than the MViTv2 model, with lower photo-z bias, scatter, and outlier fractions.  Scatter and outlier fraction are particularly sensitive to dataset size, with the MViTv2 performing much worse than the ResNet50 at small training sample sizes.  Bias appears to have no strong correlations with training set size among either model.

A full investigation into the effects of the scaling laws of these models, including pretraining dataset size, and downstream training set size (as in \cite{Walmsley24,AstroPT}) is out of the scope of this work, but merits future investigation as spectroscopic samples for many datasets, especially those from JWST, are still extremely incomplete.  While the photo-z metrics computed for the JAGUAR images are generally worse when compared to the photo-z metrics computed on the real JADES images, we focus on the relative performance to understand the scaling of each model with training set size. However, this discrepancy does point to an issue with further utilizing the simulations for our tests: there is sufficient difference between the JAGUAR and JADES images such that they cannot be combined during training to boost performance.  We have attempted to augment our JADES training data with the JAGUAR images, but no significant improvement in photo-z estimation was measured.  We match the noise level of the JAGUAR images to the real JADES observations, and limit our JAGUAR ground truth labels to objects with F444W magnitudes < 30 mag. This greatly augments the number of redshift labels used for training, and fills in more of the color space populated by the JADES spectroscopic catalog used for training.  Despite augmenting this parameter space, the addition of JAGUAR data did not help the network improve photo-z estimation on our spectroscopic test set, indicating that the domain gap between simulation and real data can be a challenge to overcome.  We discuss these points in Section \ref{sec:discussion}.  

Our fiducial model for \textsc{DeepDISC} photo-z estimation on the JWST JADES data uses a ResNet50 backbone, and is pretrained on the Galaxies ML dataset of galaxy images and spectroscopic redshifts from HSC DR2.  After training on the JADES images, we evaluate model performance using our held-out test set of spectroscopic redshifts and associated images.

\subsection{Fiducial Performance}

We quantify performance on our representative test sample of objects by examining both the redshift PDFs and derived point estimates produced from an ensemble of our fiducial model. Our ensemble is formed from a k-fold cross-validation strategy, where the data is randomly partitioned into 10 folds. Each of the 10 individual models in the ensemble uses a unique fold as a validation set, with the remaining 9 folds as a training set. Combining results from an ensemble of models helps to reduce variance in the predictions. Loss curves for these models are shown in Figure \ref{fig:ensloss}, along with the likelihood of the spectroscopic redshifts in the validation set given the the predicted photo-z PDFs.  After ~70 epochs, the validation losses appear to plateau.  To avoid overfitting, we employ an early stopping method wherein we use the model checkpoints at 70 epochs for our analysis.  

Point estimate metrics calculated for the test sample for each model in the ensemble are listed in Table \ref{tab:ens-metrics}.  We compute the bias as the median $e_z$, where $e_z = (z_{\rm spec} - z_{\rm phot})/(1+z_{\rm spec})$).  Scatter is given as $\sigma_{\rm IQR}=(e_{z75}-e_{z25})/1.349$, where the subscripts indicate percentiles of the $e_z$ distribution.  Finally, outlier fraction $\eta$ is defined as the fraction of objects with $|e_z|$>0.15.  To remove selection effects, metrics are calculated for the intersection of the set of objects detected by each model. For a final result, we combine every model's estimate by adding the each model's estimated GMM for each object.  The final photo-z PDF for each object is then given by 
\begin{equation}
    p(z) = \sum_m^{10} w_m \sum_{j=1}^5 w_{j} \mathcal{N}(z | \mu_{j}, \sigma_{ij})
\end{equation}
where $w_m$ is a weight determined from normalizing the validation set redshift likelihoods at the end of the training.  The mode of these PDFs are used as point estimates $z_{\rm phot}$ for every object. The final model yields a bias of 0.0035 ($\pm$ 0.0052), scatter of 0.0311 ($\pm$ 0.0022) and outlier fraction of 0.0503 ($\pm$ 0.0099) on a test set sample of 298 objects.  Errors are derived from the standard deviation of the ensemble results.

Our final photo-z estimates, formed from our model ensemble, are also shown in Figure \ref{fig:scatter_bins}, binned by redshift ($\Delta z=0.5$).  We color points in each bin by the F444W Kron magnitude: blue points are those brighter than the median magnitude in a given bin, and red are those fainter than (or equal to) the median magnitude.  Generally, no systematic bias is seen for either bright or faint sources.  Notable photo-z outliers appear at redshifts $\sim$0.5, $\sim$4.0 as well as $z\sim$6 and $z>9$.  We discuss these cases in the next Section. 

Detection completeness of the models can be measured with respect to the JADES DR2 catalog.  Models detect on average $\sim$70\% of the objects in our spectroscopic test set.  However, many of these objects are truncated in the creation of the sub-images, causing the network to see only a limited portion of the object light profiles. Truncations are determined based on whether a galaxy's bounding box extends beyond the edge of a sub-image. To combat this problem, we separately perform detection on cutouts re-centered on the 173 truncated objects in our spec-z test set, in order to ensure all of the pixel information for these objects is input to the network. Following this procedure, the average detection completeness with the spectroscopic test set is 95\% for models in the ensemble.  In total, 298 objects are commonly detected by all models, yielding 90.5\% detection completeness.  A few of these missing objects are very bright point sources, and the bounding boxes used to match \textsc{DeepDISC} predictions and the DR2 catalog can become very distorted due to the diffraction spikes.  Otherwise, the majority appear to have close companions of comparable or greater brightness, and thus the networks are unable to distinguish individual sources.

A comparison of our method to the photo-zs produced using \texttt{EAZY} is presented in Figure \ref{fig:EZ_comp_scatter}.  We limit the comparison to the objects detected in the test set by our model ensemble. The JADES DR2 GOODS-S \texttt{EAZY} photo-z catalog is produced using additional HST photometry where available, making the comparison not a straightforward one-to-one.  Differences in performance can thus be partly attributed to this extra optical data that \textsc{\textsc{DeepDISC}} does not have access to. The DR2 catalog generally produces better photo-zs than \textsc{DeepDISC}, with $\sigma_{\rm IQR, \texttt{EAZY}}=0.0198$ and $\sigma_{\rm IQR, \textsc{DeepDISC}}=0.0311$, and bias$_{\rm \texttt{EAZY}}=0.0023$ vs bias$_{\rm \textsc{DeepDISC}}=0.0035$.  \textsc{DeepDISC} outlier fraction is slightly better ($\eta_{\rm \textsc{DeepDISC}}=0.0503$ vs $\eta_{\rm \texttt{EAZY}}$=0.0705) although this can be attributed to the selection effect due to only including \textsc{DeepDISC}-detected sources.  We discuss this more in Section \ref{sec:discussion}.  

We also compare \textsc{DeepDISC} results to an \texttt{EAZY} run limited to the same 9 NIRCam filters. We adopt the new JADES \texttt{EAZY} templates and photometric offsets for GOODS-S field presented in \cite{Hainline24}. We follow their procedure but limit the filters to nine NIRCam filters for GOODS-S field. To distinguish these results from the \texttt{EAZY} photo-zs in the JADES GOODS-S DR2 catalog, we refer to them as ``\texttt{EAZY}9".  This result is shown in the right panel of Figure \ref{fig:EZ_comp_scatter}.  We find that the reduced number of filters leads to a degraded \texttt{EAZY} performance. When \textsc{DeepDISC} is put on even footing with \texttt{EAZY}, it is able to surpass the template-based code in terms of photo-z scatter and outlier fraction, and is comparable in bias.

We look for possible biases in our error estimates through examining $z_{\rm spec}-z_{\rm phot}$, normalized by the photo-z error derived from the 68\% credibility interval.  This error-normalized $\Delta z$ is shown in Figure \ref{fig:Norm_error_hists} and generally follows a Gaussian distribution for objects both relatively bright or faint in F444W.  JADES DR2 \texttt{EAZY} and \texttt{EAZY}9 results are qualitatively similar, although the narrower peaks suggest errors may be slightly overestimated.  We additionally examine the coverage of PDFs, i.e., fraction of PDFs that contain $z_{\rm spec}$ within an interval defined by the  credibility level, vs the credibility level.  Ideally, this value is one-to-one and along the diagonal, e.g., 68\% of PDFs contain the $z_{\rm spec}$ within their 68\% credibility interval.  In Figure \ref{fig:coverage_intervals}, we generally see good agreement with the diagonal, although there is a slight bias indicating that the \textsc{DeepDISC} errors may be too broad.   We plot the JADES DR2 \texttt{EAZY} and \texttt{EAZY}9 68 and 95 percent credibility levels for comparison.  At 68 percent credibility levels, both \texttt{EAZY} results yield a slightly larger than 68 percent coverage, further supporting overestimated errors. At 95 percent credibility levels, \texttt{EAZY} errors are either unbiased or slightly underestimated. 

\textsc{DeepDISC} is competitive with \texttt{EAZY}, and slightly outperforms it when the input filters between the two algorithms are consistent.  Its redshift posteriors appear to be well-calibrated and unbiased in terms of object brightness.  We reiterate that a main advantage of \textsc{DeepDISC} is the speed ease of use - one only needs to input imaging data after training, which cuts down on computational costs due to source extraction and aperture photometry, which are necessary for template-fitting photo-z codes.  However, in order to use \textsc{DeepDISC} to the fullest potential, one must be aware of its behavior in different circumstances, and the caveats therein.  In the following, we critically examine the behavior of the network.

\begin{table}
    \centering
    \begin{tabular}{crrr}
        \hline
        \hline
         & Bias $<e_z>$ & scatter $\sigma_{\rm IQR}$ & outlier fraction $\eta$ \\ 
         \hline
        Combined &  0.0035 &  0.0311 & 0.0503 \\
        \hline
        Fold 0 &  0.0129 &  0.0370 & 0.0772 \\
        Fold 1 &  0.0132 &  0.0388 & 0.0839 \\
        Fold 2 &  0.0056 &  0.0356 & 0.0604 \\
        Fold 3 &  0.0071 &  0.0410 & 0.0839 \\
        Fold 4 &  0.0095 &  0.0420 & 0.0738 \\
        Fold 5 &  0.0052 &  0.0391 & 0.0570 \\
        Fold 6 &  -0.0057 &  0.0379 & 0.0839 \\
        Fold 7 &  0.0031 &  0.0391 & 0.0638 \\
        Fold 8 &  0.0030 &  0.0435 & 0.0738 \\
        Fold 9 &  0.0041 &  0.0409 & 0.0839 \\
        \hline
    \end{tabular}
    \centering
    \caption{Point statistics for 10 models, each trained on one of the training data folds. The combined row indicates statistics computed from the mode produced by adding the GMMs from model.  Statistics are computed on the intersection of objects detected by each model. Variations between models can be attributed to selection effects.}
    \label{tab:ens-metrics}
\end{table}

\section{Discussion}
\label{sec:discussion}

We first examine the \textsc{DeepDISC} behavior due to choices of input filters.  Notably, the degeneracy at $z\sim4$ and $z\sim0.5$ is not as strongly present in \textsc{DeepDISC} photo-zs.  The Lyman break at high redshift can appear similar to the Balmer break at low redshift, leading to confused photo-z estimates which can be further complicated by emission lines.  However, the Lyman break is not visible in the bluest NIRCam filter (F090W) until $z\sim9$, so the network may not be as susceptible to this degeneracy compared to the JADES DR2 \texttt{EAZY} photo-zs which use supplementary HST imaging in bluer filters. \texttt{EAZY}9 results still yield a number of outliers at these redshifts, indicating that pixel-level information helps disentangle .  Higher scatter in the \textsc{\textsc{DeepDISC}} and \texttt{EAZY}9 photo-zs at $z\sim 5-6$ may be due to the Balmer break shifting to the short and long wavelength filter dichroic cutoffs.

\begin{figure*}
    \centering
    \includegraphics[width=0.66\textwidth]{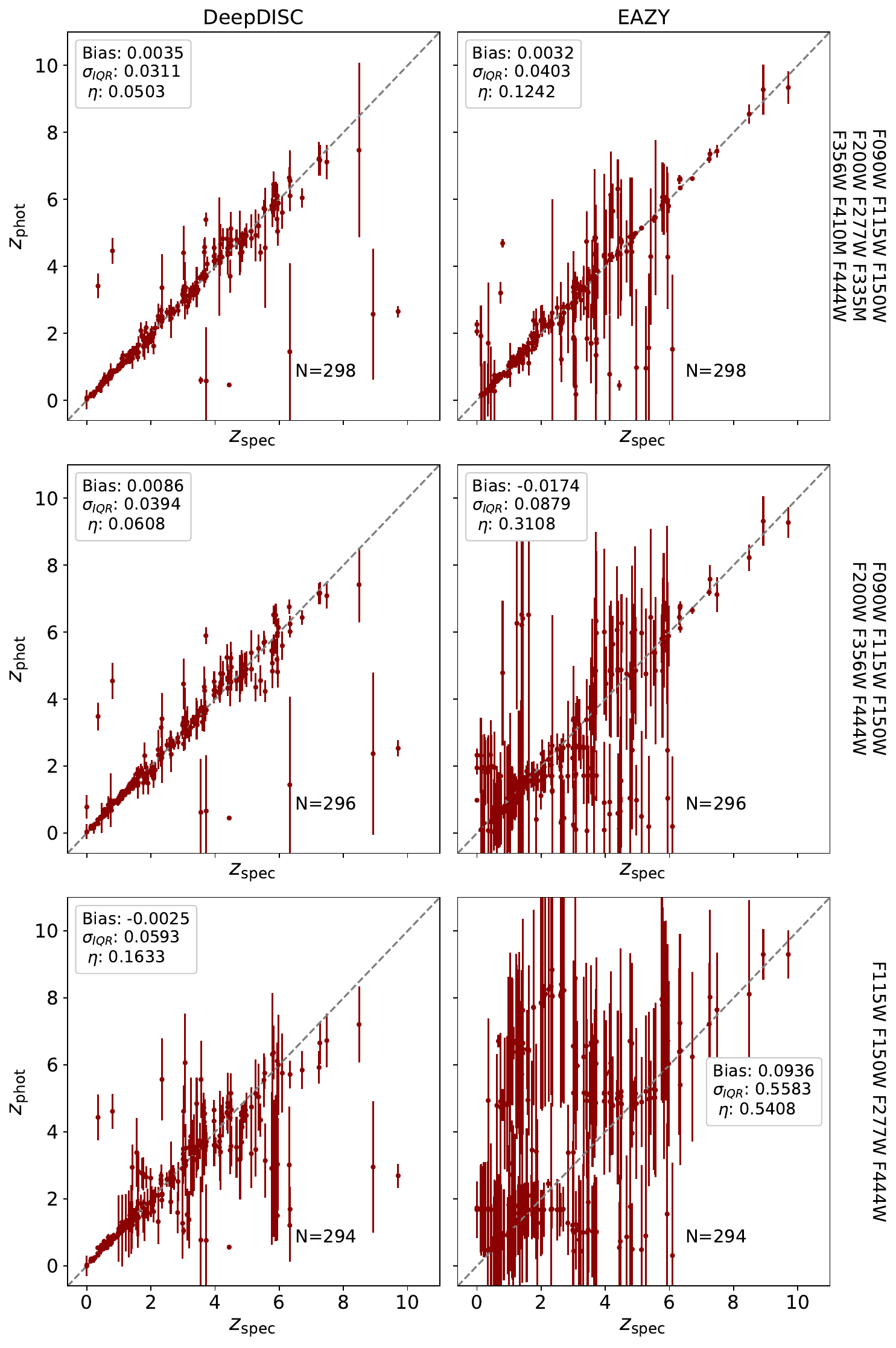}
    \caption{$z_{\rm spec}$ vs $z_{\rm phot}$ for JADES (top) NEXUS (middle) and COSMOS-Web (bottom) filters.  The left column shows the \textsc{DeepDISC} estimates, and the right shows \texttt{EAZY} estimates.  Error bars are derived from the 68\% confidence intervals of the photo-z PDFs.  The number of detected objects and the filter combinations are included for reference.  \textsc{DeepDISC} yields better point estimates compared to \texttt{EAZY} using the same set of filters, and is more robust to a smaller number of input filters than \texttt{EAZY}.}
    \label{fig:surveys_scatter}
\end{figure*}

We repeat the previous analysis with different subsets of the filters to gauge the impact on photo-z performance. We choose the NIRCam filter configurations used in the COSMOS-Web imaging \citep[F115W, F150W, F277W and F444W;][]{COSMOSWeb} and the North ecliptic pole EXtragalactic Unified Survey  \citep[F090W, F115W, F150W, F200W, F356W, and F444W;][NEXUS]{NEXUS, NEXUS-EDR}. We perform the same steps: creating a model ensemble, training until convergence, and combining ensemble results into a single set of redshift PDFs for the test set.  Photo-z point estimates are shown in Figure \ref{fig:surveys_scatter}, with errors derived from 68\% credibility levels.  Photo-z performance generally worsens with a smaller number of available filters, as expected.   Photo-z errors visibly increase as filters are removed, reflecting the larger uncertainties when less color information is provided.  In all cases, \textsc{DeepDISC} photo-zs are significantly better than their \texttt{EAZY} counterparts, indicating that pixel-level information is useful across a large range of redshift space.

\begin{figure}[htb!]
    \centering
    \includegraphics[width=\columnwidth]{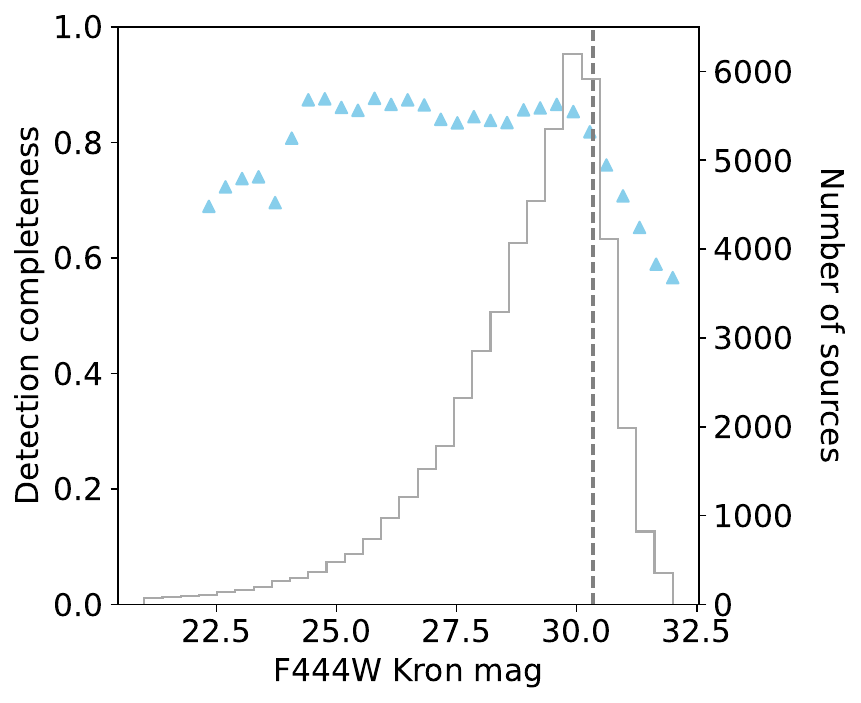}
    \caption{Detection completeness (blue triangles) of \textsc{DeepDISC} as function of F444W Kron mag.  The sample only consists of objects with imaging in the 9 NIRCam filters used during training. The histogram shows the F444W Kron mag distribution of objects in the sample.  The dashed line is at the $5\sigma$ detection limit for F444W.  Average completeness is at around 90\%, dropping off sharply at the edges of the mag distribution. }
    \label{fig:dectcomp}
\end{figure}

\begin{figure*}
    \centering
    \subfloat{\includegraphics[width=0.33\textwidth]{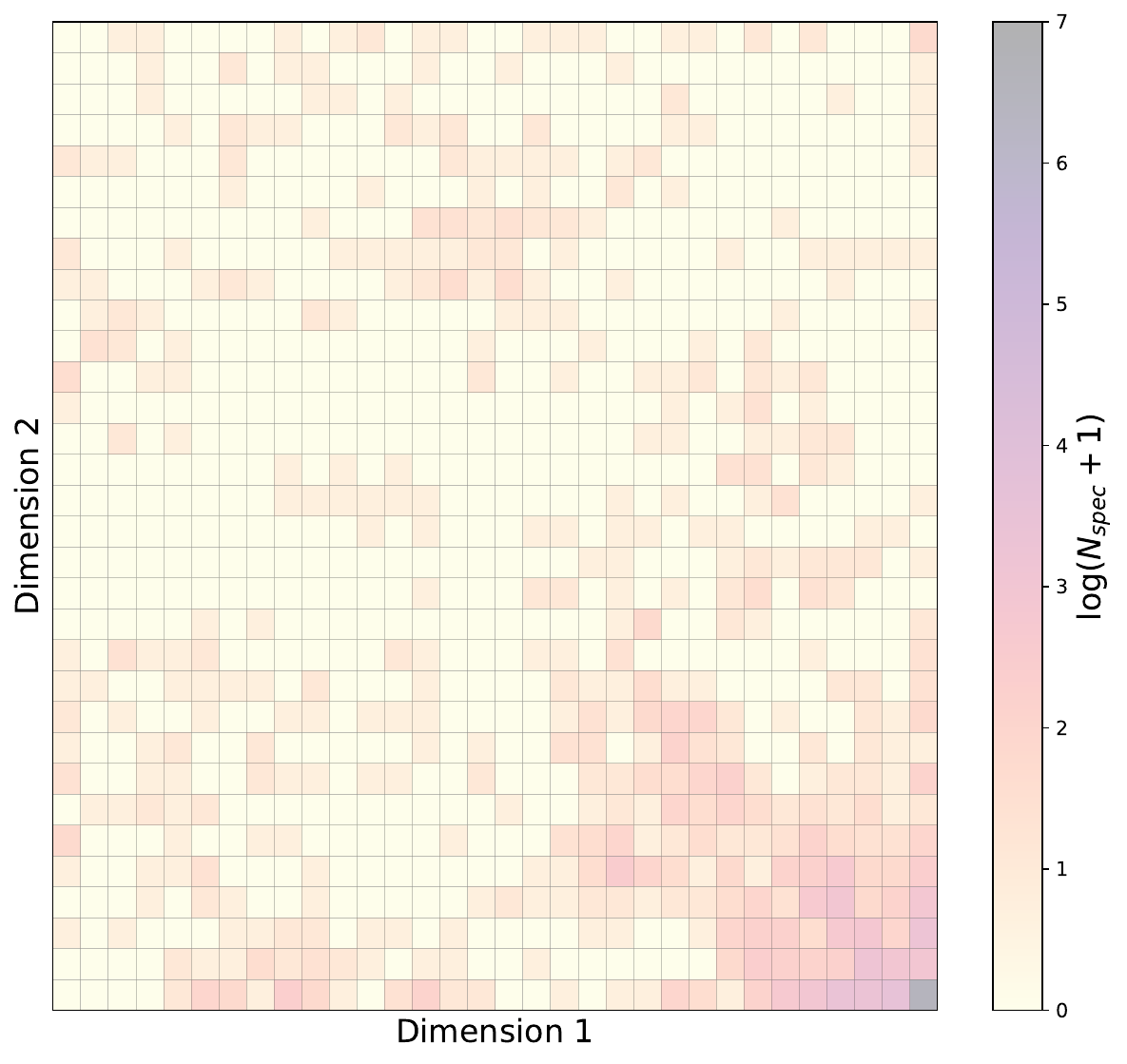}}
    \subfloat{\includegraphics[width=0.33\textwidth]{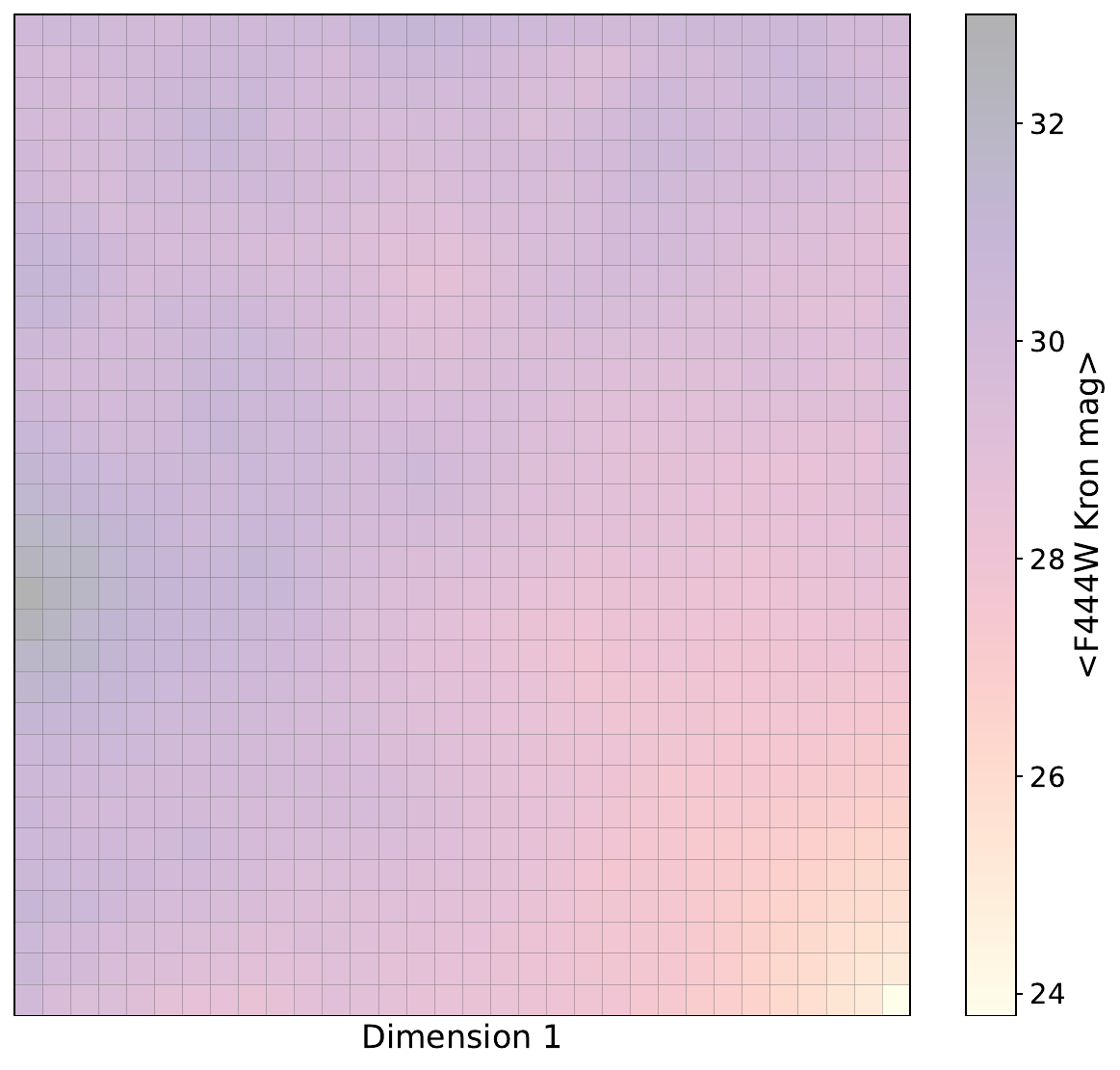}}
    \subfloat{\includegraphics[width=0.33\textwidth]{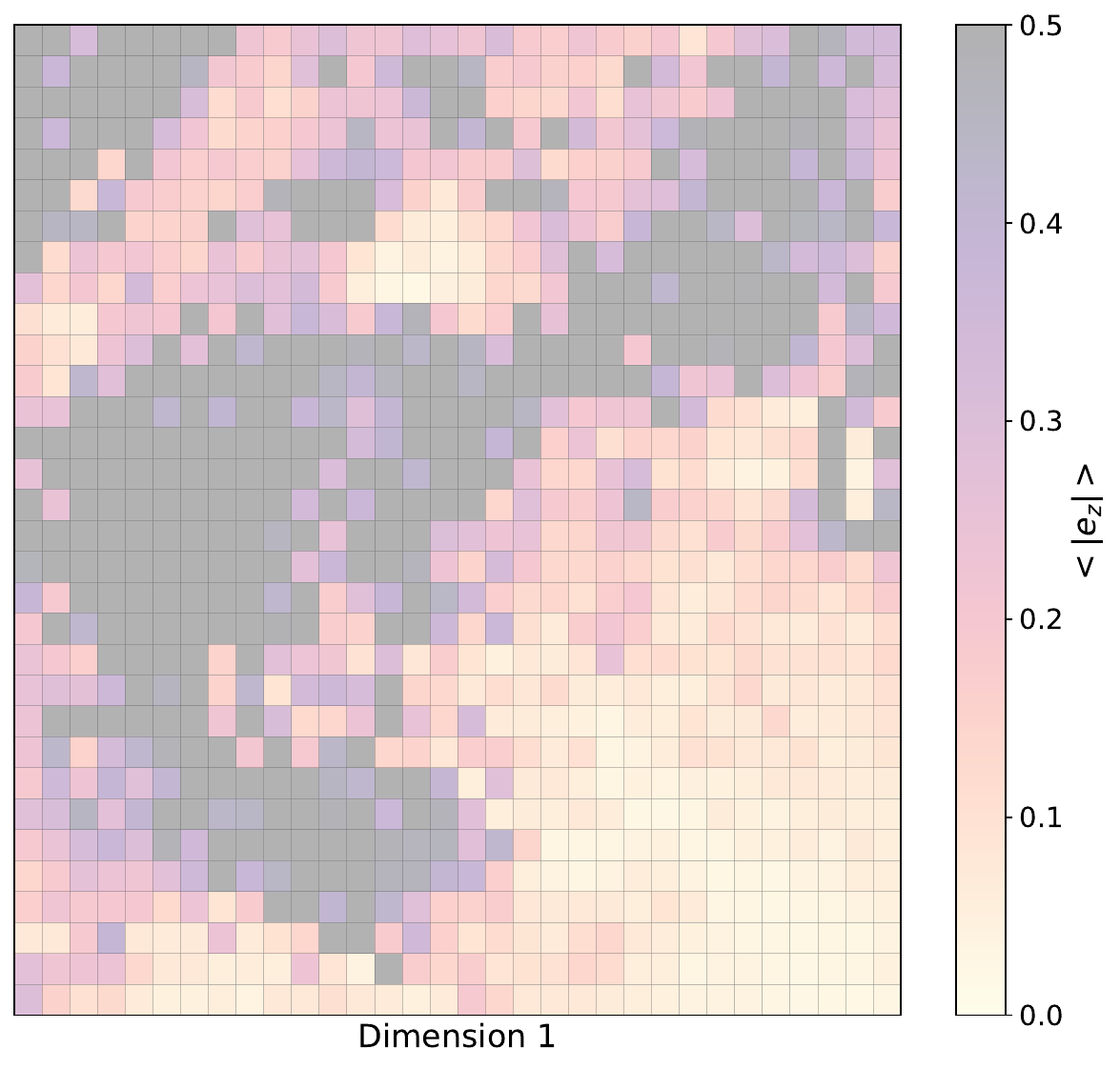}}
    \caption{SOM cells derived from the colors of objects in the JADES DR2 GOODS-S catalog with imaging in the 9 JADES NIRCam filters. Colors indicate per-cell quantities of the number of training sample spectroscopic redshifts per cell (left), median F444W Kron magnitude (middle) and median absolute bias $<|e_z|>$ between \textsc{DeepDISC} and JADES DR2 \texttt{EAZY} (right). Color bar min/max values are truncated to aid visibility.  Many regions of the SOM space are either completely devoid of any  training data, or sparsely populated.  In general, photo-zs for objects that reside in such cells should be considered untrustworthy.  We note, however, that some underrepresented cells yield reasonable photo-z estimates based on our test set of spectroscopic redshifts and the JADES DR2 \texttt{EAZY} catalog.}
    \label{fig:soms}
\end{figure*}

\subsection{Applications to the full JADES photometric sample}

Having validated our model on a representative test set, we now apply it to the full JADES GOODS-S photometric sample and discuss practical considerations.  In order to produce photo-z estimates for the entire JADES GOODS-S catalog, we first divide the JADES imaging into 20x20 equal sized blocks. Then, we create sub-images with a size equal to the block size, and with a 500 pixel overlap with neighboring sub-images.  This ensures we minimize any effects due to objects being truncated at the edges of sub-images.  We then run the images through our ensemble of trained models.  Each model detects, segments, and produces a photo-z estimate for objects in the images.  We combine ensemble predictions by taking the intersection of objects that were detected by each model.  The resulting \textsc{DeepDISC} catalog is 80\% complete with the JADES DR2 GOODS-S catalog for objects with imaging in all 9 NIRCam filters. This is a lower overall detection completeness compared to that with our spectroscopic test sample, which we attribute largely to the fact that the spectroscopic test sample is biased to be brighter than the rest of the catalog.  Our detection completeness is sensitive to object magnitude, and drops steeply towards the 5$\sigma$ detection limit (Figure \ref{fig:dectcomp}). However, some brighter objects are still missed, which often have bright close-by companions.  It is unclear why the network misses these objects, but possible remedies include tuning hyperparameters that control the detection/blending sensitivity or altering the image pixel scaling to a quasi-logarithmic scale that highlights the extended regions of objects \citep{Merz2023}.

To provide a fully complete photo-z catalog, we run \textsc{DeepDISC} with "forced photometry mode" on the objects in the JADES DR2 GOODS-S catalog that were not detected by the ensemble. This amounts to fixing the input bounding boxes, rather than bounding box predictions being carried out by the network. We provide a \textit{forced} flag, indicating that a photo-z was produced with forced photometry mode.

Our previous comparisons with \texttt{EAZY} in Section \ref{sec:results} only included objects that \textsc{DeepDISC} detected (298 out of 330 in our test set).  We recompute photo-z metrics for our entire test sample using the complete photo-z catalog, i.e., with the 32 additional sources added from "forced photometry mode".  The results are shown in Table \ref{tab:comp-metrics-withforced}, compared with JADES DR2 \texttt{EAZY} and \texttt{EAZY9}.  All metrics degrade for all methods, although \textsc{DeepDISC} still outperforms \texttt{EAZY9}. The \textsc{DeepDISC} outlier fraction now exceeds that of JADES DR2 \texttt{EAZY} . There are a few stellar contaminants in the additional sources, but the majority are faint galaxies.  Excluding the stellar sources, the average F444W Kron magnitude of the additional objects is 26.75 mag, compared to 25.18 mag for the original 298 objects.

\begin{table}
    \centering
    \begin{tabular}{crrr}
        \hline
        \hline
         & Bias $<e_z>$ & scatter $\sigma_{\rm IQR}$ & outlier fraction $\eta$ \\ 
         \hline
        DeepDISC &  0.0022 &  0.0378 & 0.0970 \\
        EAZY DR2 &  0.0024&  0.0222 & 0.0879 \\
        EAZY9 &  0.0040 &  0.0449 & 0.1576 \\
        \hline
    \end{tabular}
    \centering
    \caption{Photo-z point estimate metrics including objects originally not detected by DeepDISC.}
    \label{tab:comp-metrics-withforced}
\end{table}

\subsection{Training Representation}
Our results in Section \ref{sec:results} have shown the viability of photometric redshift estimation via deep learning for JWST images.  \textsc{DeepDISC} is able to push to redshifts $z\sim8$, and produces better estimates than \texttt{EAZY} if the same photometric filters are used.  However, a major disadvantage of the network is the challenge of application to a broader sample of objects that may not be represented in the training data.  This is an inherent problem with all machine learning methods, as they can struggle to make predictions in regions of feature space that they have not sufficiently learned. \cite{Moskowitz24} and \cite{Zhang25} address this problem with data augmentation, generating new data that fills in regions of color/magnitude space that were underrepresented in the training data.  However, these works focus on catalog-based machine learning techniques that rely on aperture photometry rather than raw pixel data. Pixel based data augmentation is much more challenging due to the higher dimensionality.  Generative models \citep{Lanusse21,smith_realistic_2022} may be a way to address this problem for image-based estimators, but their adoption lies outside the scope of this work.  As discussed in the previous Sections, we attempted to augment our training data with simulated images of JADES galaxies, but found no notable improvement in photo-z estimation.

Due to the aforementioned issues with unrepresented regions in color/magnitude space during training, we produce a \textit{spec-rep} flag indicating if the object is not well-represented by the training set in color/magnitude space.  To produce this flag, we construct a self-organizing map (SOM) using the DR2 GOODS-S photometric catalog.  A SOM is a popular dimensionality reduction technique due to its ease of visualization and interpretability.  It assigns objects to a 2D grid of cells based on a learned mapping from the input features.  We create our SOM using the \textsc{minisom} python package.  We take the set of JADES DR2 GOODS-S objects with imaging in all 9 NIRCam filters, and input the F444W magnitudes, along with sequential colors (e.g., F115W-F090W, F150W-F115W, etc.). We use principal component analysis (PCA) weight initialization and choose a 32x32 SOM with rectangular topology, gaussian neighborhood function with $\sigma=5$ and learning rate 0.5.  The resulting SOM cells are shown in Figure \ref{fig:soms}, colored by the number of spectroscopic training set objects in a given cell, as well as median F444W Kron magnitude and the average $|e_z|$ between \textsc{DeepDISC} and JADES DR2 \texttt{EAZY} estimates.  The cell with the most training sample spec-zs corresponds to the cell with the highest median F444W Kron mag, reflecting the fact that the spectroscopic selection favors bright objects.  The SOM is included in our data release. 

The \textit{spec-rep} flag represents the number of spec-zs that populate a given object's SOM cell and can be used to select photo-zs based on training sample representation (higher=more trustworthy).  50.3\% of the DR2 objects with 9 NIRCam filters do not have any spectroscopic representation in their corresponding SOM cell (corresponding to 72.7\%) of all objects.  The relative dearth of spectroscopic data will be a persistent challenge for broad applicability of machine learning photo-z methods.  We note that there are some test set objects with spectroscopic redshifts that are assigned to SOM cells without training set representation.  By looking at this subset, we can gain some insight into how much the training sample representation in the SOM matters for photo-z estimations.  We find that while there are 9 photo-z outliers in this subset of 32, the remaining photo-zs generally have good agreement with the spectroscopic redshifts.  This indicates that the SOM does not fully describe the calibration of \textsc{DeepDISC} photo-zs, based on spectroscopic representation.  An interplay of a few factors contributes to this: (1) The hyperparameters of the SOM.  Increasing the number of cells would increase the expressive power of the SOM at the expense of capturing more noise.  Decreasing the cell count would cause individual cells to contain a larger volume of color space.  There are different methods for choosing the number of cells \citep[e.g.,][]{Masters15, Zhang25} and these choices are influenced by the science cases that require photo-z calibration. (2) Deep learning methods may be able to interpolate between regions of color-redshift space.  Recent work \citep{Moran26} has shown that pixel-based methods are able to produce better photo-z estimates within regions of color space, i.e., within the same (relatively large) SOM cell, compared to catalog-based machine learning methods.  The authors attribute this to the fact that deep learning algorithms reduce attenuation bias (bias towards the mean redshift in a SOM cell) by potentially learning an optimal extraction of information from pixels over integrated aperture photometry.
The combination of these factors indicate that there may be regions of our SOM-derived color space that, despite having no training sample spec-zs for calibration, still correspond to reliable \textsc{DeepDISC} photo-zs.  Conversely, a single training set spec-z in a SOM cell may not be enough data for \textsc{DeepDISC} to adequately estimate photo-zs.  To provide some insight into this, we compare the left and right panels of Figure \ref{fig:soms}.  The left panel shows the training sample representation in the SOM and the right panel shows the average bias $|e_z|$ between \textsc{DeepDISC} and the JADES DR2 \texttt{EAZY} catalog.  Given the lack of spectroscopic redshifts for the whole sample, we use the JADES DR2 \texttt{EAZY} catalog as a rough proxy for $z_{\rm spec}$ to set a baseline. We see that there are some regions of the SOM that have training sample representation but are very biased compared to JADES DR2 \texttt{EAZY} photo-zs.  On the other hand, there are some SOM cells that have no training sample representation and yet are at percent level bias with JADES DR2 \texttt{EAZY}.  An in-depth investigation into understanding and optimizing training sample calibration for deep learning photo-z methods would be of great benefit for future studies.

\subsection{Future Perspectives}

Although this work presents a significant first step in applying deep learning learning methods to JWST data for photometric redshift estimation, there is clearly room for improvement.  We have explored different model architectures and pretraining setups, gaining insights into strengths and weaknesses of various schemes.  One of the major limiting factors in our method (and in any machine learning method) is the quantity and quality of the training data.  JWST data, while of extremely high quality due to its high resolution, wavelength coverage, and lack of atmospheric contributions to the PSF, is still quite limited in volume.  While we have shown that neural networks can learn useful features for photo-z estimation if they are pretrained on images from an entirely different telescope, extrapolation to regimes underrepresented by the training data is still unwise.  Simulated data can be used to augment regions of color/magnitude space in real data that do not have spectroscopic measurements, but our attempts to use JAGUAR image simulations did not yield any significant improvements. A possible solution to this domain gap lies in using adversarial training, as in \citep{Huertas-Company23}.  Other simulations may prove to be more useful in transferring knowledge to real data, such as IllustrisTNG \citep{TNG1,TNG2,TNG3}, which has been used to calibrate the morphological classification of JWST CEERS images with deep learning \cite{VF24}.  Unlabeled data could be leveraged using contrastive learning to ensure that the networks produce features that are more strongly correlated with redshift.  

The strong performance of \textsc{DeepDISC} for bright sources, even with limited wavelength coverage, suggests that the model could be effective for shallow or pure-parallel programs, where morphology and spatially resolved color information remain informative despite reduced depth.  \cite{Merz25} find that \textsc{DeepDISC} can derive photo-zs with less than <1\% bias and scatter for bright objects (<20 mag) with 6 filters sampling rest frame UV to optical ($\sim$0.3-1.1 $\mu$m) while also observing that the photo-z scatter is roughly inversely proportional to the signal-to-noise of the images.  We therefore expect similar qualitative behavior for JWST and other datasets. A dedicated, quantitative exploration of depth-dependent performance, e.g., via controlled noise injection or depth-matched training with noise augmentations would help to further characterize model performance.

A potentially difficult regime for \textsc{DeepDISC} is crowded field imaging of galaxy clusters. Blending contamination from intracluster light and bright galaxies, as well as the distortion of source morphologies due to strong lensing make photometry in these crowded fields particularly challenging \citep{Jouvel14, Connor17}. While the detection aspect of our model may struggle on crowded fields (indicated by the detection incompleteness reported above), we expect \textsc{DeepDISC} photo-zs to be in general less sensitive to blending compared to methods that use deblended catalog photometry, as \cite{Merz25} demonstrate with simulated LSST data.  The high resolution of JWST compared to LSST gives us a cautionary optimism for performance in this regime.  Given the data-driven nature of \textsc{DeepDISC}, an adequate training sample of cluster images including strongly lensed source galaxies would likely enable robust detection and photo-z estimation of both lens and source galaxies.  Simulation packages such \textsc{skylens} \citep{Plazas19} could produce simulated training data, although our experience with the JAGUAR images indicates that the simulation-to-real domain transfer should be carefully treated.  For application to galaxy clusters, adding source/lens classifications to \textsc{DeepDISC} could help the network with photo-z estimation since this establishes a somewhat informative prior (sources need to be at higher redshift than lens galaxies).
Finally, as discussed above, pixel-based methods may provide a unique advantage with strongly lensed galaxies, due to their hypothesized ability to learn an optimal weighting of individual pixels that could overcome the limitations of integrated photometry \citep{Moran26}.

Our code is open source for users who want to train and use their own models for custom datasets.  We provide tutorials and documentation for data preprocessing, training and model evaluation.

\section{Conclusions}

\textsc{DeepDISC} is a holistic neural network framework designed to detect, deblend, and classify sources using multiband imaging as input.  The framework has been extended to perform photometric redshift estimation, and in this work we have pushed the capabilities of the network to high-z regimes with JWST data.  We benchmark multiple scenarios – comparing different model  architectures, testing the effect of pretraining on terrestrial vs galaxy images, and assess performance under different filter sets (JADES 9-filter set vs subset combinations like COSMOS-Web with 4 filters and NEXUS with 6 filters). This comprehensive analysis provides insights into the model behavior, and directions for future work.

Our main results are:
\begin{itemize}
    \item \textsc{DeepDISC} can produce accurate photo-z estimates on JWST images up to $z\sim8$.  The spectroscopic training sample defines regions in color/magnitude space where our photo-zs are trustworthy. 
    \item When input filters are the same, \textsc{DeepDISC} outperforms \texttt{EAZY}, in terms of point estimate scatter and outlier fraction.  Errors produced by the PDFs appear to be mostly unbiased, but perhaps slightly overestimated.
    \item Unsurprisingly, model performance degrades and error estimates increase as the number of photometric filters decreases.  \textsc{DeepDISC} estimates are more robust to fewer filters than \texttt{EAZY}. Our model's performance compared to \texttt{EAZY} is significantly better with NEXUS (6 NIRCam filters) and COSMOS-Web (4 NIRCam filters) filter configurations. This result highlights \textsc{DeepDISC}'s advantage over template fitting for photo-z estimation in surveys with limited filter combinations.  
    \item The transformer model previously used as a baseline for \textsc{DeepDISC} is not optimal for small data regimes of only a few thousand training labels, e.g., a limited sample of JWST data.  This is not unexpected, due to previous studies on the behavior of transformers with increasing label size.
    \item A ResNet model can outperform the transformer baseline if pretrained on galaxy images and spec-zs.  Pretraining on galaxy images allows for the ResNet model to produce features that are more informative for redshifts compared to pretraining with ImageNet or training from scratch.  It is not necessary for the pretraining dataset to be JWST imaging, indicating that astronomy foundation models could be used within the \textsc{DeepDISC} framework to infer photo-zs.
    \item Our model performance scales with training set size, indicating that there is room for improvement with more data.  Data augmentation with generative models could be used to fill in gaps of color space that are un-or under-represented by the training data.
    
\end{itemize}

Photo-z estimation with deep learning has become an active area of research as computer vision models mature and data volumes increase.  In low and mid range redshift regimes, image-based deep learning has been shown to generally outperform methods that rely on aperture photometry.  With JWST and the upcoming Nancy Grace Roman Space Telescope, high redshift regimes are being observed, providing key insights into the early universe.  Photo-zs remain a necessary ingredient in studies utilizing these data.  Template-based methods remain dominant in the literature, but as machine learning methods evolve, it is worth exploring their application.  The main limitation of machine/deep learning methods is sparse training data which can be especially limiting for near-infrared, space-based imaging. However, the body of data keeps growing, with programs like ASTRODEEP \citep{Merlin2024}, NEXUS and COSMOS catalogs \citep{COSMOS-specz}. New datasets will be a valuable resource for further training, as photo-z estimates will improve and the application sample of objects will increase.  Deep learning has already produced valuable insights from JWST data, and we believe that much more lies ahead.  This work provides groundwork for new avenues of exploration and synthesizes insights from the broader computer vision literature, while also clearly outlining limitations and necessary areas of caution.  We are optimistic for future applications, and welcome continued testing and use of \textsc{DeepDISC} by the broader community. 

\label{sec:conclusions}

\section*{Acknowledgements}

We thank the anonymous referee for helpful comments. We thank Alex Malz for helpful discussion with TheLastMetric and pzflow. 
We thank Dr. S. Luo at the National Center for Supercomputing Applications (NCSA) for helpful discussion and assistance with the GPU cluster used in this work. 
G.M. thanks the LSST-DA Data Science Fellowship Program, which is funded by LSST-DA, the Brinson Foundation, and the Moore Foundation; his participation in the program has benefited this work.
G.M. and X.L. acknowledge support from Illinois Campus Research Board Award RB25035, NSF grant AST-2308174, and NASA grants 80NSSC24K0219 and 80NSSC26K0333.

This work is based in part on observations made with the NASA/ESA/CSA James Webb Space Telescope. The data were obtained from the Mikulski Archive for Space Telescopes at the Space Telescope Science Institute, which is operated by the Association of Universities for Research in Astronomy, Inc., under NASA contract NAS 5-03127 for JWST. These observations are associated with programs 1180,
1210, 1895, 1963, 3215, and 4540.  The authors acknowledge the teams of programs 1895, 1963, and 3215 for developing their observing program with a zero-exclusive-access period.

This work utilizes resources supported by the National Science Foundation's Major Research Instrumentation program, grant \#1725729, as well as the University of Illinois at Urbana-Champaign. 
This work used Delta at NCSA through allocation PHY240290 from the Advanced Cyberinfrastructure Coordination Ecosystem: Services \& Support (ACCESS) program, which is supported by U.S. National Science Foundation grants \#2138259, \#2138286, \#2138307, \#2137603, and \#2138296.

We acknowledge use of \texttt{Matplotlib} \citep{Hunter2007}, a community-developed Python library for plotting. This research made use of \texttt{Astropy},\footnote{\href{http://www.astropy.org}{http://www.astropy.org}} a community-developed core Python package for Astronomy \citep{astropy:2013, astropy:2018}, \texttt{Pandas} \citep{mckinney-proc-scipy-2010} and \texttt{SciPy} \citep{2020SciPy-NMeth}.  This research has made use of NASA's Astrophysics Data System.

This research has made use of the NASA/IPAC Infrared Science Archive, which is funded by the National Aeronautics and Space Administration and operated by the California Institute of Technology.

\section{Data Availability}
The photo-z catalog produced by our fiducial ensemble for the JADES DR2 GOODS-S sample and the SOM used in the analysis are available at \href{https://doi.org/10.5281/zenodo.17487691}{https://doi.org/10.5281/zenodo.17487691}.  The \textsc{DeepDISC} code is publicly available at \href{https://github.com/grantmerz/deepdisc}{https://github.com/grantmerz/deepdisc}, along with tutorials.




\bibliographystyle{mnras}
\bibliography{ref} 




\appendix

\section{Mutual Information}
\label{app:MI}

\begin{figure*}
    \centering
    \includegraphics[width=0.8\textwidth]{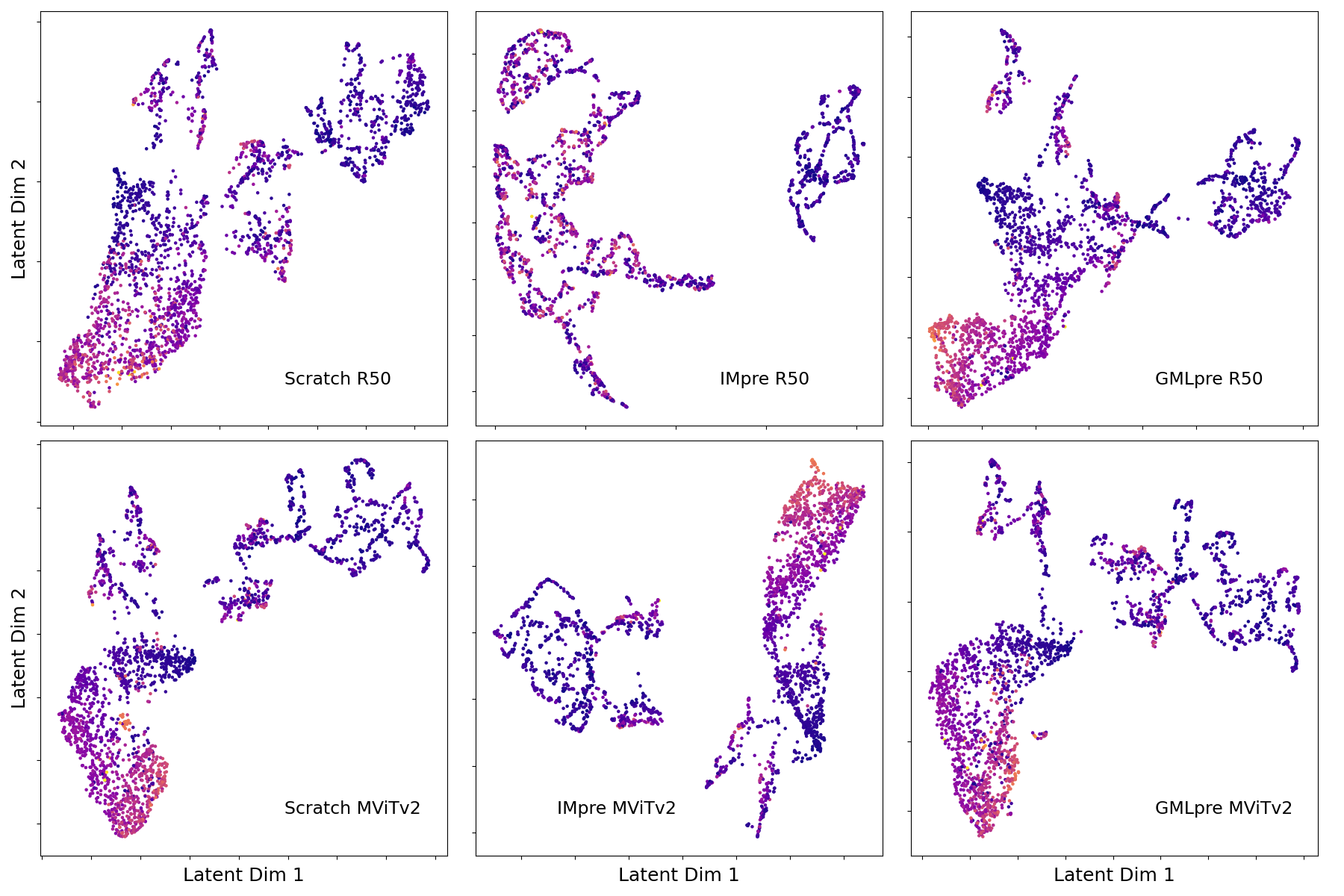}
    \caption{UMAP distribution of \textsc{DeepDISC} features, reduced to two dimensions.  Different architectures and pretraining schemes are shown in the different panels.  Points are colored by redshift.  Training and test set objects are included.}
    \label{fig:UMAP}
\end{figure*}

\begin{figure*}
    \centering
    \subfloat{\includegraphics[width=0.3\textwidth]{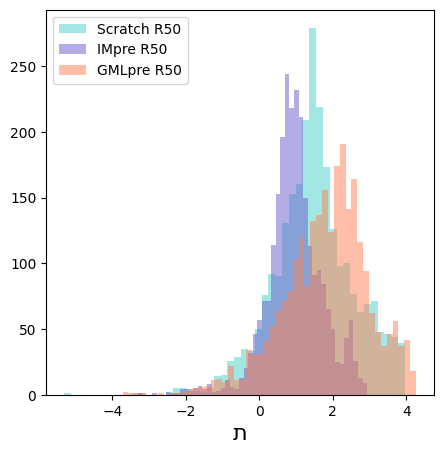}}
    \subfloat{\includegraphics[width=0.3\textwidth]{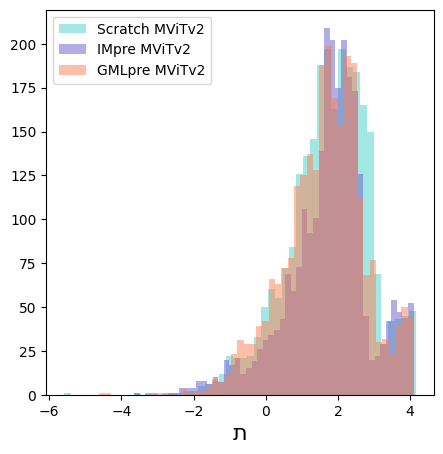}}
    \caption{$\tav$, a lower bound on mutual information between redshift and neural network features.  Tav is computed using a normalizing flow to approximate the conditional distribution of features and redshifts.  Tav is calculated pre-object for training and test set objects and binned. ResNet50 results are shown on the left, and MViTv2 on the left.  Different pretraining schemes are shown in different colors. }
    \label{fig:tavs}
\end{figure*}

We qualitatively examine the ability of the networks to extract information relevant to redshift through Uniform Manifold Approximation Projection \citep[UMAP][]{mcinnes2018umap-software} visualizations.  We show how features extracted by the networks depend on redshift in Figure \ref{fig:UMAP}.  After training, we run extract features from the JADES data by fixing the input bounding boxes to those made by the DR2 analysis.  This is analogous to running forced photometry, and we do this to avoid any effects due to sample selections.  Features are then reduced to 2 dimensions using a UMAP reduction.  We extract features from both the training and test sets, given the small sample size of JADES images.  We assume that the ability of the networks to extract useful information on the training set aligns with their ability to extract information on unseen test sets.  We also test training a network from scratch, i.e., no pretraining, for reference.  Figure \ref{fig:UMAP} shows that the all models have some level of structure that correlates with redshift, i.e., color gradients in the UMAP, except for the ImageNet-pretrained ResNet50 model. 

Quantitatively, we compute an approximation for the mutual information between model features and redshift.  Mutual information measures how much uncertainty of a measurement of a random variable decreases due to knowing another random variable.  Mutual information is a useful metric due to its ability to capture any dependencies among the random variables.  It has been used to elucidate neural network latent spaces \citep{GMMMI}, examine how neural networks compress data vs generalize information \citep{Goldfeld18}, and conduct self-supervised training \citep{Hayat21}. \cite{Scott25} used mutual information between photometry and redshift to assess how overlapping astronomical surveys could benefit photometric redshift estimates. Comparisons of MI between model features and redshift can be used to see which model extracts the most useful information for photo-z estimation.  We use TheLastMetric, defined in \cite{TLM}, which approximates a lower bound on the mutual information between random variables $x$ and $y$

\begin{equation}
    \tav = E_{p(x,y)}log\, q_\phi(x|y) + H(x)
\end{equation}
where $q_\phi(x|y)$ is an estimate of the true distribution $p(x|y)$ and $H(x)$ is the entropy of $p(x)$.  This estimate is done using a normalizing flow trained with a Kullback-Leibler divergence (see \cite{TLM}) for details.  We train the normalizing flow using pzflow \citep{Crenshaw24}.  In our case, $x$ is the redshift and $y$ is the set of features for each object.  We reduce the features to 11 dimensions using a UMAP to help with convergence of the flows.  The flows require a bijector to map the input to a compact latent space before training.  We use a rolling spline coupling to map our catalog of 11 dimensional features to a uniform latent space.  We train 30 flows in an ensemble, each for 200 epochs and with a learning rate of 1e-4, reduced by a factor of 10 every 50 epochs.  After training, we use the flows to compute $\tav$ for every object in our input data sample.  The distributions of $\tav$ for each model/pretraining scheme are shown in Figure \ref{fig:tavs}.  The Figure shows that pretraining largely has no effect on the ability of MViTv2 models to extract information relative to redshift, as $\tav$ distributions are similar across all pretraining schemes.  However, the ResNet50 model is sensitive to pretraining, with clearly distinct $\tav$ distributions for each pretraining scheme. There is a gain in mutual information between neural network features and redshift when the GML dataset is used to pretrain our model.  Despite the different properties of the GML and JADES data, it appears that giving the network prior information about the relationship between galaxy images and redshift helps decode that relationship on new data.  This prior is strong for a ResNet, and weak for an MViTv2 model.  This can be explained in part by the weaker inductive biases inherent in transformers, due to their attention mechanism over convolutions.  The MViTv2 model generalizes well to new datasets (as in \cite{Merz2023}) but does not extract the most useful information compared to an "in-domain" pretrained ResNet.  A big factor in this discrepancy is likely the training sample size.  Our JADES training set contains 914 images and 1937 unique objects with a spectroscopic redshift.  Despite the high resolution and rich color information, this small data set poses a challenge to transformer models, which are known to struggle in small data regimes.

\label{lastpage}
\end{document}